\documentclass[a4paper, 12pt]{article}
\usepackage[a4paper,margin=1in]{geometry}
\usepackage[round,sort]{natbib}
\usepackage{hyperref}
\usepackage{float}
\usepackage{graphicx}
\usepackage{dcolumn}
\usepackage{csquotes}
\usepackage{amsmath}
\usepackage[utf8]{inputenc}
\usepackage[swedish, english]{babel}
\usepackage{caption}
\usepackage{subcaption}
\usepackage{tikz}
\usepackage{rotating}

\usetikzlibrary{arrows,chains,matrix,positioning,scopes}
\makeatletter
\tikzset{join/.code=\tikzset{after node path={\ifx\tikzchainprevious\pgfutil@empty\else(\tikzchainprevious)edge[every join]#1(\tikzchaincurrent)\fi}}}
\makeatother
\tikzset{>=stealth',every on chain/.append style={join},
every join/.style={->}}
\tikzstyle{labeled}=[execute at begin node=$\scriptstyle,
execute at end node=$]

\begin{document}

\title{\textbf{A study protocol for a comparative effectiveness evaluation of
antiandrogenic medications against Standard of Care}}
\author{Paulina Jon\'{e}us\hspace{0.2cm} \\
Department of Statistics, Uppsala University,\\
Per Johansson \\
Department of Statistics, Uppsala University, IFAU and Tsinghua University\\
Sophie Langenskiold,  Department of Medical Sciences, Cardiology }
\maketitle

\begin{abstract}
This paper presents a study protocol for an effectiveness evaluation of antiandrogenic (NAM) medications (i.e. abiraterone acetate and enzalutamide) against standard of care in patients suffering from metastatic castrate resistant prostate cancer. The study protocol takes stock in the Rubin Causal Model for observational data in using historical comparisons for the analysis on three outcomes: mortality, pain and skeleton related events. We restrict the evaluation to 1,285 NAM patients with a prostate cancer diagnosis after June 2012 and find the comparisons in the group of men that had a prostate cancer diagnosis between June 2008 and June 2010. The rich registry data and the lack of NAMs as an option for the comparisons provide support for the validity of the design. However, while the design yields balance on the observed covariates, one cannot discard the possibility that unobserved confounders are not balanced. Thus, sensitivity analyses are suggested. \end{abstract}

\section{Introduction}
This paper presents a study protocol for a comparative effectiveness evaluation of antiandrogenic medications (NAM) versus standard of care (SoC) in patients suffering from metastatic castrate resistant prostate cancer (mCRPC). The NAMs are abiraterone acetate (AA) in combination with prednisone and enzalutamide (ENZ). The protocol define the treatment and control groups and specify the analyses to be conducted before outcome data has been accessed, just like in randomized controlled trials. 

Before June, 2015, almost no patients were prescribed any of the two NAMs as they were not reimbursed. The drugs started to be used after June 15 2015 when the Dental and Pharmaceutical Benefit Agency (TLV) reimbursed AA for mCRPC patients who had failed docetaxel. In July 2015 TLV also reimbursed prescriptions of ENZ for the same group of patients. In June 15 2018 TLV reimbursed AA and ENZ for patients who had failed androgen deprivation therapy (ADT) but were not yet suited for docetaxel.  

Results presented in \cite{Johansson_et_al_2021a} show that docetaxel is the standard of care (SoC) against which the NAM should be compared.  As docetaxel was the SoC for this group of patients since it was approved in 2004, and since almost no patient was prescribed a NAM before June 15 2015, it is reasonable to assume that controls can be found using historical data. 

The data is taken from population registers administrated by the National Board of Health and Welfare (NBHW) and Statistics Sweden (SCB). As a consequence of including another population to be reimbursed in June 15 2018 we restrict the population of NAM patients to those prescribed AA or ENZ during the period June 1 2015 to June 15 2018. However, to restrict the comparisons from being prescribed a NAM in any given follow up period, there is a need to select a sub-population of NAM patients. We sampled $1,285$ men diagnosed with a prostate cancer (PC) before June 1 2012 and who were prescribed a NAM within 36 months after the diagnosis. The comparison group was selected from the pool of all men who had a PC diagnosis between June 1 2008 and June 1 2010, leaving us with $19,456$ patients.\footnote{Details on the sampling and the strategy of selecting controls from the comparisons is given in section \ref{sec:Sample}.} These sample restrictions allow the follow-up time to be between two and five years. 

The advantage of focusing on the sub-population of NAM patients with a fast progression, i.e. prescribed a NAM within 36 months after the diagnosis, is that the two populations can be sampled close in time. This means that both the demography of the two groups and the health care offered should be quite similar, except for the differences in the two treatments. The drawback of using this sub-population is that the analysis will be restricted to a group of patients with a more severe disease progression than in the overall population. 

The maintained assumption of the study design is that given observed pre-treatment covariates, the treatments (NAM or SoC) are unconfounded. This means that we observe the relevant covariates that are determining the treatments and that are associated with the potential outcomes. According to \cite{Johansson_et_al_2021a}, disease progression is an important determinant for the choice between NAM and SoC. Thus, a valid study design needs to control for covariates during the disease progression. We, thus, include covariates measured at the time of diagnosis and for each month after the diagnosis, up to the month prescribed the NAM. For the comparisons we create 36 covariate \enquote*{trajectories}. Each trajectory summarizes the disease progression in any of the 36 months following the PC diagnosis. A SoC patient with \enquote*{the same} covariate trajectory as the NAM patients would be considered as a good control to the NAM treated. As we are using the entropy balancing scheme of \cite{hainmueller2012entropy}, a control means, loosely, having a large weight in contrast to having a small or zero weight. The weights in the entropy balancing are derived from imposing balance constraints directly on the covariates of the NAM treated and controls. The analysis on the three outcomes; mortality, pain and skeleton related events will be conducted using weighted least squares estimation.

While the design yields balance on the observed covariates, one cannot discard the possibility that unobserved confounders are not balanced. For this reason, a sensitivity analysis is suggested. To this end, we will be using data from the national prostate cancer register (NPCR) that will be added after publishing the protocol. The sensitivity analysis is a placebo test, that is, we estimate effects where there should not be effects if the design is valid. The placebo test will make use of three covariates measure before NAM prescription. These covariates, judged, by specialist to be potential confounders, are the PSA levels, the Gleason score, and metastases. If there is a statically significant effect on these covariates, this suggests that available data from population registers is not sufficient to control for confounding.

The resulting analysis should be of large importance as prostate cancer is
the second most commonly diagnosed solid organ malignancy in the world. In
Sweden, it constitutes 30\% of all cancers diagnosed in the male population, which corresponded to 10,500 new cases in 2016. PC is also the fifth leading cause of cancer death worldwide and second leading cause of cancer mortality among males in the USA \citep{rawla19}, and almost all mortalities arise when the patients have progressed to mCRPC.

The rest of the paper has the following structure. Section \ref{sec:Data}
describes the data and discusses in details the specific sampling scheme of design and the construction of 
covariates. Section \ref{sec:Ident} provides the assumption underlying the
design together with an analysis of the common support assumption and details on the entropy balancing scheme. The protocol for the analyses is given in section \ref{sec:Plan}. The paper ends with a discussion in section \ref{sec:Discussion}.

\section{Data and sample restrictions}
\label{sec:Data}

The data was collected from linked population registers administrated by the National Board of Health and Welfare (NBHW), and Statistics Sweden (SCB). All in- and outpatient care visits in Sweden and all prescribed drugs are registered in the registers from NBHW. The inpatient care register contains among others information on all diagnoses (using the ICD10 classifications), the date of admission, and discharge. Based on this information we restrict the population to all men in Sweden with a PC diagnosis (i.e. ICD10 C61.9) in the period 1986 to 2016. The pharmaceutical register contains the date of prescription and dispensing of drugs, and also, the ATC class of the drug. 

The SCB data includes information on age, marital status, educational level, and country of birth as well as a large set of covariates measuring pensions, income, sick leave and other security benefits for the patient and the household. We collect data on socio-economic status at the year of diagnosis and the two preceding years. In a few cases of missing values, the average of the existing information from previous years is used for each individual. Information on educational level is sometimes missing. In these cases, a five nearest neighbor approach is used to impute missing values. That is, the most common value of educational level of the five individuals, i.e., neighbors, who are most similar when it comes to income, pension, age and country of birth, is imputed.
\subsection{Sample restrictions}

\label{sec:Sample}

We have access to the date of PC diagnosis for everyone in our data, but we do not observe when patients have been administered SoC. The fact that the two drugs were not offered by the Swedish health care before being subsidized means that defining a comparison group of historical patients could be a valid strategy in the evaluation. However, the new drugs were not restricted only to new diagnoses, this needs to be considered in defining the comparison population. 

The comparison group should be as similar as possible to the NAM group, and the two groups should have been provided the same quality of health care. If the comparison group is sampled close in time to the NAM group, these two restrictions may hold. The comparison group needs, however, to be sampled in a manner such that the outcome is evaluated before a potential prescription of a NAM. 

As an illustration of the problem, consider a patient diagnosed with a PC in June 2010. This patient can be given a NAM in June 2015 after a duration to be prescribed (DTP) a NAM of five years. Therefore, the population of NAM treated is restricted to the population who have had a diagnosis in the period June 2012 to June 2015 and with a DTP of less than 37 months. Comparisons, sampled in June 2010 have a total follow up time of five years before, if still alive, being offered a NAM. This means that they can be used as comparisons to all prescribed a NAM at all possible DTP in the restricted population. In order to obtain a sufficiently large group of comparison patients, we sample all men with PC diagnosis in the period June 1 2008 to June 1 2010, which gives us a sample of $19,456$ patients. As we will have mortality data to June 2020, we have a minimum follow up period of two years and a maximum follow up period of five years. 

\begin{figure}[ht]
\centerline{
\makebox[\textwidth][c]{\includegraphics[width=.8\textwidth]{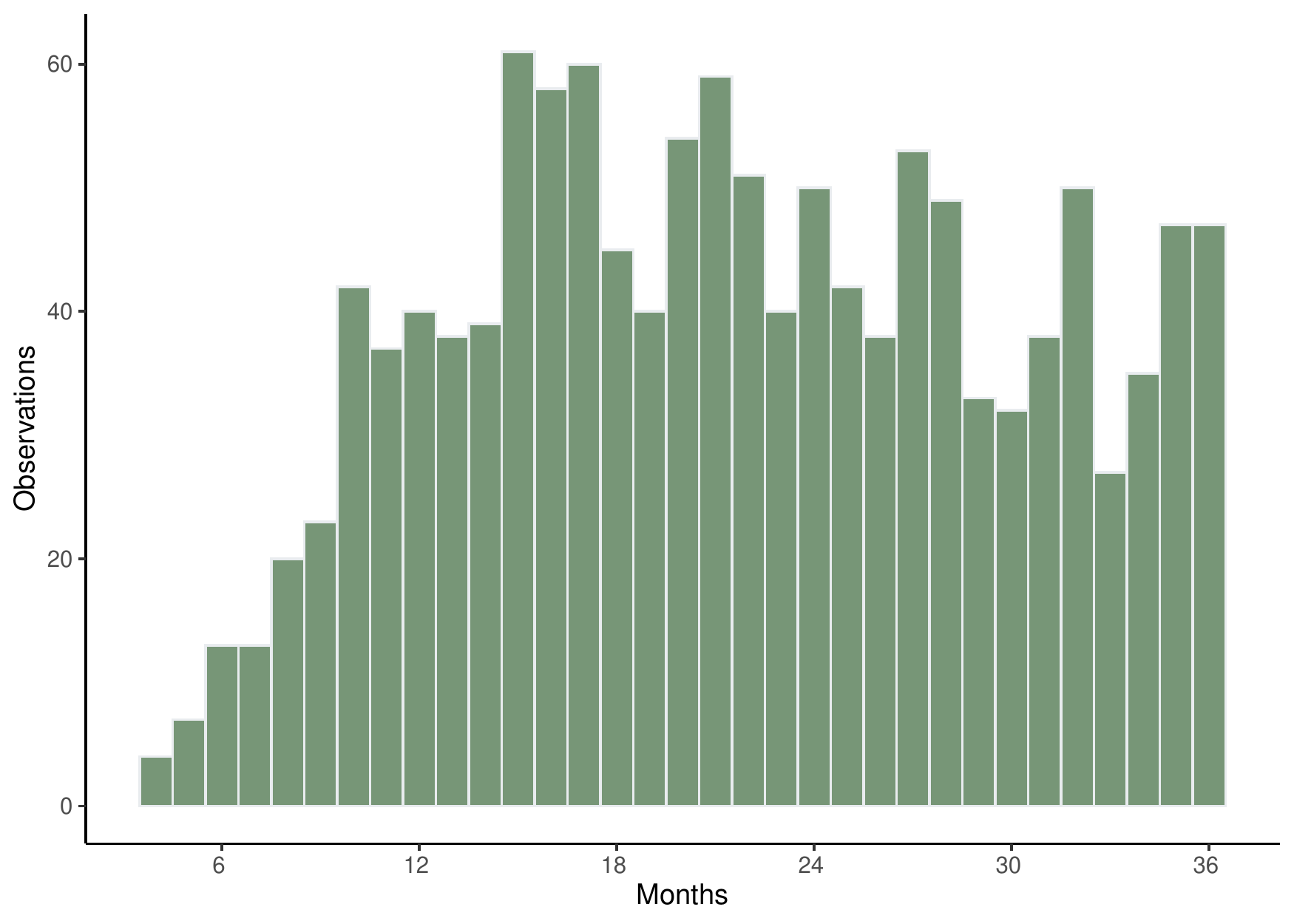}} 
}
\caption{Distribution of time between diagnosis and NAM prescription.}
\label{DISTR}
\end{figure}

We restrict the evaluation to 1,285 NAM patients with a prostate cancer diagnosis in June 2012 to June 2015 and who were prescribed a NAM within 36 months after the diagnosis. The distribution of months until NAM treatment is presented in Figure \ref{DISTR}. From this figure we can see that a few patients were prescribed a NAM within 6 months after the diagnosis, but the majority was prescribed a NAM after more than 12 months.

The assignment of treatment is the core in any empirical study. Results in \cite{Johansson_et_al_2021a} show that the general state of health is important. Being of old age, presence of comorbidity or suffering from generally poor health were signals to be prescribed ENZ or AA. The progression of metastases is also deemed as important. A fast progression indicated docetaxel as the preferred first line treatment irrespective of the patient's health. As health progression is an important determinant for prescription of a NAM, a valid study design needs to control for the health progression after diagnosis. 

Thus, for each month, we create covariates describing monthly health of the patients. For the NAM patients this is done up to the date of prescription of a NAM, and for the comparisons, health data is created for all periods up until 36 months after the diagnosis. The month of SoC treatment is defined as the treatment month of the NAM treated patient most similar to this comparison patient in terms of these monthly covariates and a large set of pre-diagnosis covariates. 

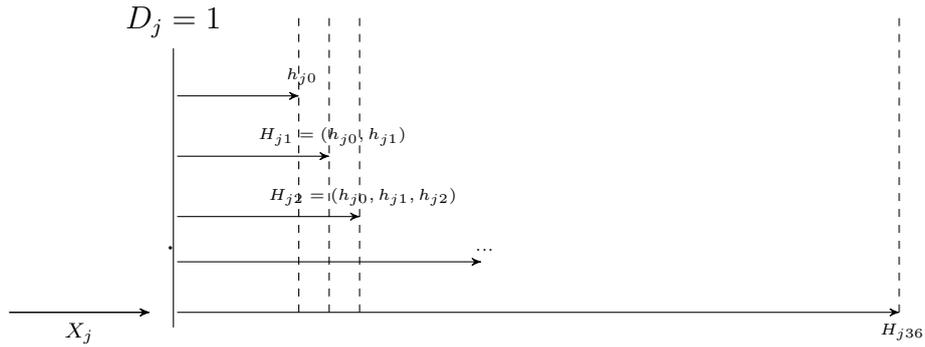
\begin{figure}[ht]
\centering
\begin{tikzpicture}
\path (-28:-1.8cm) node(p2) {$\cdot$};
\draw (-1.55,-0.2) -- (-1.55,3.5) node(yline1)[above] {$D_j=1$};
\draw[dashed, -] (0.1,- 0.0) -- (0.1,4);
\draw[->] (-1.5, 2.87) -- (0.1, 2.87)
node(yline1)[above] {\begin{tiny}
$h_{j0}$
\end{tiny}};
\draw[dashed, -] (0.5,- 0.0) -- (0.5,4);
\draw[->] (-1.5, 2.07) -- (0.5, 2.07)
node(yline1)[above] {\begin{tiny}
$H_{j1}=(h_{j0},h_{j1})$
\end{tiny}};
\draw[dashed, -] (0.9,- 0.0) -- (0.9,4);
\draw[->] (-1.5, 1.27) -- (0.9, 1.27)
node(yline1)[above] {\begin{tiny}
$H_{j2}=(h_{j0},h_{j1},h_{j2})$
\end{tiny}};
\draw[->] (-1.5, 0.67) -- (2.5, 0.67)
node(yline1)[above] {\begin{tiny}
$...$
\end{tiny}};
\draw[dashed, -] (8,- 0.0) -- (8,4);
\draw[->] (-1.5, 0) -- (8, 0)
node(yline1)[below] {\begin{tiny}
$H_{j36}$
\end{tiny}};
\matrix (m) [matrix of math nodes, row sep=6em, column sep=4.5em]
{ \text{ } & \text{ } & & & \\};
{ [start chain] 
\chainin (m-1-1);
\chainin (m-1-2) [join={node[below,labeled] {X_j}}];
}; 
\end{tikzpicture}
\caption{One comparison individual with 36 possible control observations, one for each trajectory $t=1,...,36$.}
\label{Fig:illustration}
\end{figure}

\begin{figure}[ht]
\begin{subfigure}{.5\textwidth}
\centering
\begin{tikzpicture}
\path (-28:-1.8cm) node(p2) {};
\draw[dashed, -] (.8,-0) -- (.8,3) node(yline)[above] {\textit{June 2010}};
\draw[dashed, -] (-2.5,-0) -- (-2.5,3) node(yline)[above] {\textit{June 2008}};
\draw[dashed, -] (3,-0) -- (3,3) node(yline)[above] {\textit{June 2015}};
\draw (1.65,-0.1) -- (1.65,1.5) node(yline)[above] {$T_j=0$};
\draw (-0.2,0.8) -- (-0.2,0.8) node(yline2)[above] {$w_j=w_i$}; 
\draw (-1.68,-0.1) -- (-1.68,1.5) node(yline1)[above] {$D_j=1$};
\draw[-] (-1.68, 0.87) -- (1.65, 0.87);
\draw[dashed,,->] (1.5, 0) -- (2.90, 0);
\draw[,-] (-2, 0) -- (1.65, 0);
\matrix (m) [matrix of math nodes, row sep=5.em, column sep=3.5em]
{ \text{ } & \text{ } & & \text{ } & \\};
{ [start chain] 
\chainin (m-1-1);
\chainin (m-1-2) [join={node[below,labeled] {X_j}}];
\chainin (m-1-4) [join={node[below,labeled] {H_{jw_i}=(h_{j0},...,h_{jw_i-1})}}];
}; 
\end{tikzpicture}
\caption{Historical comparison}
\label{one_hist}
\end{subfigure}
\begin{subfigure}{.5\textwidth}
\begin{tikzpicture}
\path (-28:-1.8cm) node(p2) {};
\draw[dashed, -] (.8,-0) -- (.8,3) node(yline)[above] {\textit{June 2015}};
\draw[dashed, -] (-2.5,-0) -- (-2.5,3) node(yline)[above] {\textit{June 2012}};
\draw[dashed, -] (3,-0) -- (3,3) node(yline)[above] {\textit{June 2020}};
\draw (1.65,-0.1) -- (1.65,1.5) node(yline)[above] {$T_j=1$};
\draw (-0.2,0.8) -- (-0.2,0.8) node(yline2)[above] {$w_i$}; 
\draw (-1.68,-0.1) -- (-1.68,1.5) node(yline1)[above] {$D_j=1$};
\draw[-] (-1.68, 0.87) -- (1.65, 0.87);
\draw[dashed,,->] (1.5, 0) -- (2.90, 0);
\draw[,-] (-2, 0) -- (1.65, 0);
\matrix (m) [matrix of math nodes, row sep=5.em, column sep=3.5em]
{ \text{ } & \text{ } & & \text{ } & \\};
{ [start chain] 
\chainin (m-1-1);
\chainin (m-1-2) [join={node[below,labeled] {X_i}}];
\chainin (m-1-4) [join={node[below,labeled] {H_{iw_i}=(h_{i0},...,h_{iw_i-1})}}];
}; 
\end{tikzpicture}
\caption{NAM Treated}
\label{two_hist}
\end{subfigure}
\caption{Description of how the control group is constructed. A NAM treated $i$ (right panel) and a comparison patient $j$ (left panel) are assumed to have similar covariates on socioeconomic status and health at the time of diagnosis ($X_{i}=X_{j})$ and the same health progression up to the month of treatment, i.e. $H_{iw_i}=H_{jw_i}.$ This comparison will be a control and assumed to be given a SoC at a duration $w_i$. Thus, the outcome for this patient will be taken from month $w_i+1$.}
\label{Fig:Idea}
\end{figure}
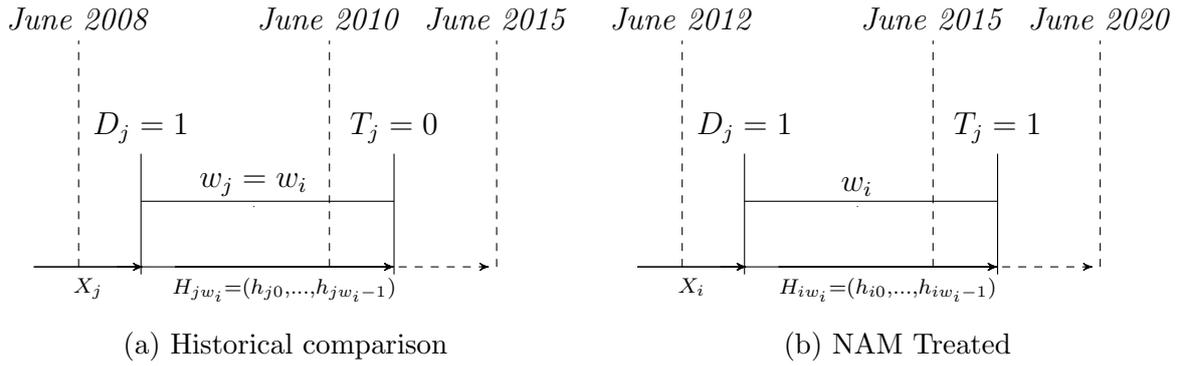

In order to provide a better understanding of the sampling design of the controls from the group of comparisons, a graphical illustration is provided in Figures \ref{Fig:illustration} and \ref{Fig:Idea}. Figure \ref{Fig:illustration} presents how the data is constructed for a comparison patient, $j$ while Figure \ref{Fig:Idea} describes the sampling of this comparison given the NAM patient $i$ displayed in the right panel of this figure. 

In Figure \ref{Fig:illustration}, $X_{j}$ denotes observed measures of socioeconomic status and health before diagnosis and $h_{jt-1}$ denote observed health measures at each month, $t=1,...,36$, after the diagnosis. The health progression up to month $t$ is denoted as $H_{it}=(h_{i0}, h_{i1},... h_{it-1})$. Furthermore, $D_i=1$ if an individual is diagnosed with PC (zero else) and we let $T_i=1$ if patient $i$ is prescribed a NAM (zero else). 

The NAM patient $i$ displayed in Figure \ref{two_hist} is prescribed a NAM at month $w_{i}$ and he is observed to have had a health progression up to $w_{i}$ of $H_{iw_i-1}$. If the comparison individual displayed in the left panel are similar on  $X_{i}$ and $H_{iw_i-1}$ we assume patient $j$ would have been given a SoC treatment (i.e. $T_j=0$) instead of a NAM (as this was not an option) at DTP $w_{j}=w_{i}.$ 

\begin{sidewaysfigure}
\centering
\includegraphics[scale=.8]{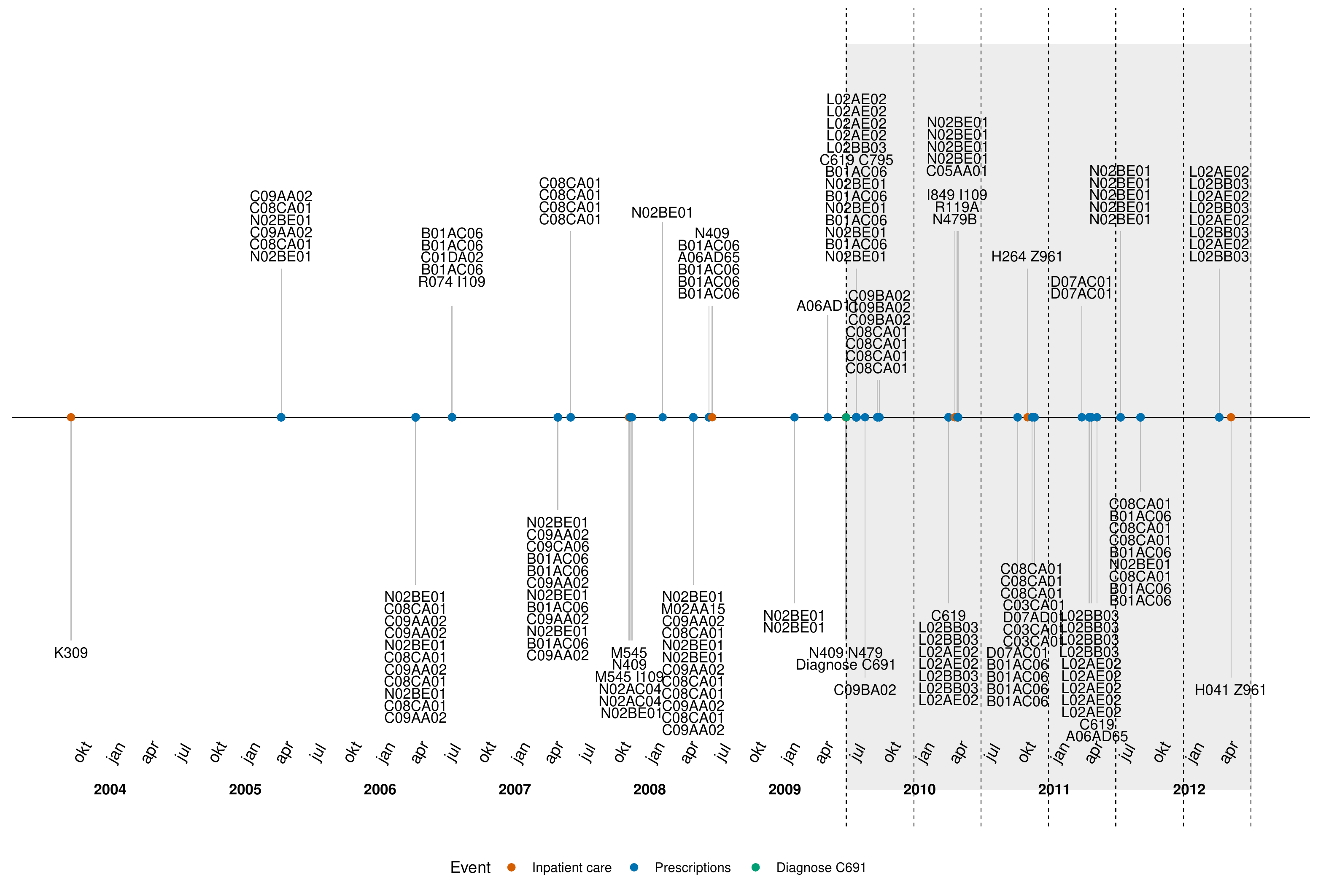} 
\caption{One chosen control individual}
\label{Fig:Control_ind}
\end{sidewaysfigure}

\begin{sidewaysfigure}
\centering
\includegraphics[scale=.8]{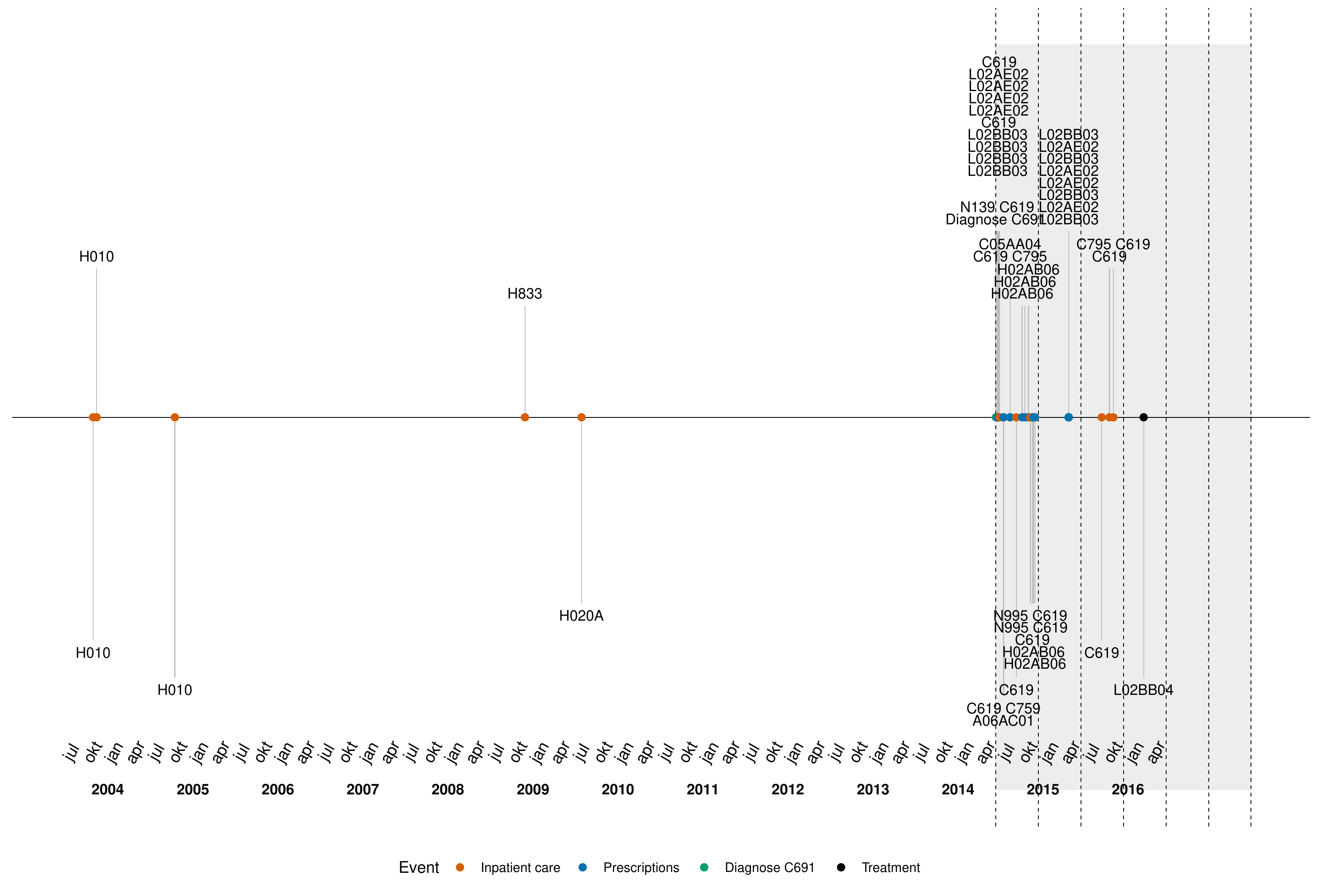} 
\caption{One chosen treatment individual}
\label{Fig:Treat_ind}
\end{sidewaysfigure}

Figures \ref{Fig:Control_ind} and \ref{Fig:Treat_ind} illustrate the level of details in the available health data by showing all collected drugs and health care visits of a comparison and NAM treated patient, respectively. The shaded areas represent three years after being diagnosed with PC. 

Figure \ref{Fig:Control_ind} reveals that the comparison patient was of age 78 when he was diagnosed with PC in 2009. It can be noted that he was treated with both bicalutamide and a GnRH-analog (ATC codes L02BB03 and L02AE02) and that he had a skeleton cancer diagnose (ICD C795) in relation to the PC diagnose. The NAM treated patient in Figure \ref{Fig:Treat_ind} was diagnosed with PC in June 2014 and was diagnosed for skeleton metastases two months later. After being treated with both bicalutamide and a GnRH-analog he was treated with ENZ 21 months later. 

\subsubsection{Pre-diagnosis covariates}
From the NBHW registers we calculate the
Elixhouser comorbidity index.\footnote{%
To this end the R-package \href{https://cran.r-project.org/web/packages/comorbidity/vignettes/comorbidityscores.html}%
{\textit{Comorbidity }}is used.} The number of hospital visits 1, 1-6, 6-12 and
1-60 months before diagnosis are also included to account for the health status of the patient before the diagnosis.

From the SCB data we include age, marital status (married or not married),
country of birth (born in the nordic countries or not born in the nordic
conuntries) and educational level. Educational level is the highest
completed education and is classified as less than, equal or more than
secondary school education. 

We reduce the dimension of 42 continuous covariates on socio-economic status measured before diagnosis by using an exploratory factor analysis and add five factors. Looking at the total sample, the results from the factor analysis are provided in Table \ref{Factormodel}. The first 5 factors explain 59 percent of the variation of these covariates. The first factor loads mainly on pension payments (private, old age pension and payment from negotiated agreements), the second on social assistance, the third loads on payments from early retirements, the forth loads on capital
income, income from work and disposable income. The fifth loads on income from work
and disposable income.

Descriptive statistics on pre-diagnosis covariates across the two groups are displayed in Table \ref{ttest}. From the table we can see that those prescribed NAM is older and that their health is worse than that of the comparison group. Looking at number of health visits between 6 and 60 months before the diagnosis, there is no statistically significant health difference between the two groups. However, in the periods just before the diagnosis we find substantial differences in the number of health visits. Even though the fraction with worst co-morbidity index (index $\geq 5)$ is quite similar across the two groups, it is quite clear that the NAM patients have worse health than the comparisons also with regard to the co-morbidity index. The reason is that the fraction with the best health (index group $0$) is higher for the comparison group than for those prescribed NAM. Interestingly, as it is in line with the hospital visits, the difference between the co-morbidity indices is smaller when measured 12 months before the diagnosis than at the time of the diagnosis. All in all, there is clear indication of more severe health progression of the NAM treatment group than in the comparison group. The differences in socioeconomic status do not seem to be substantial.

\begin{figure}[tbp]
\centering
\includegraphics[scale=.6]{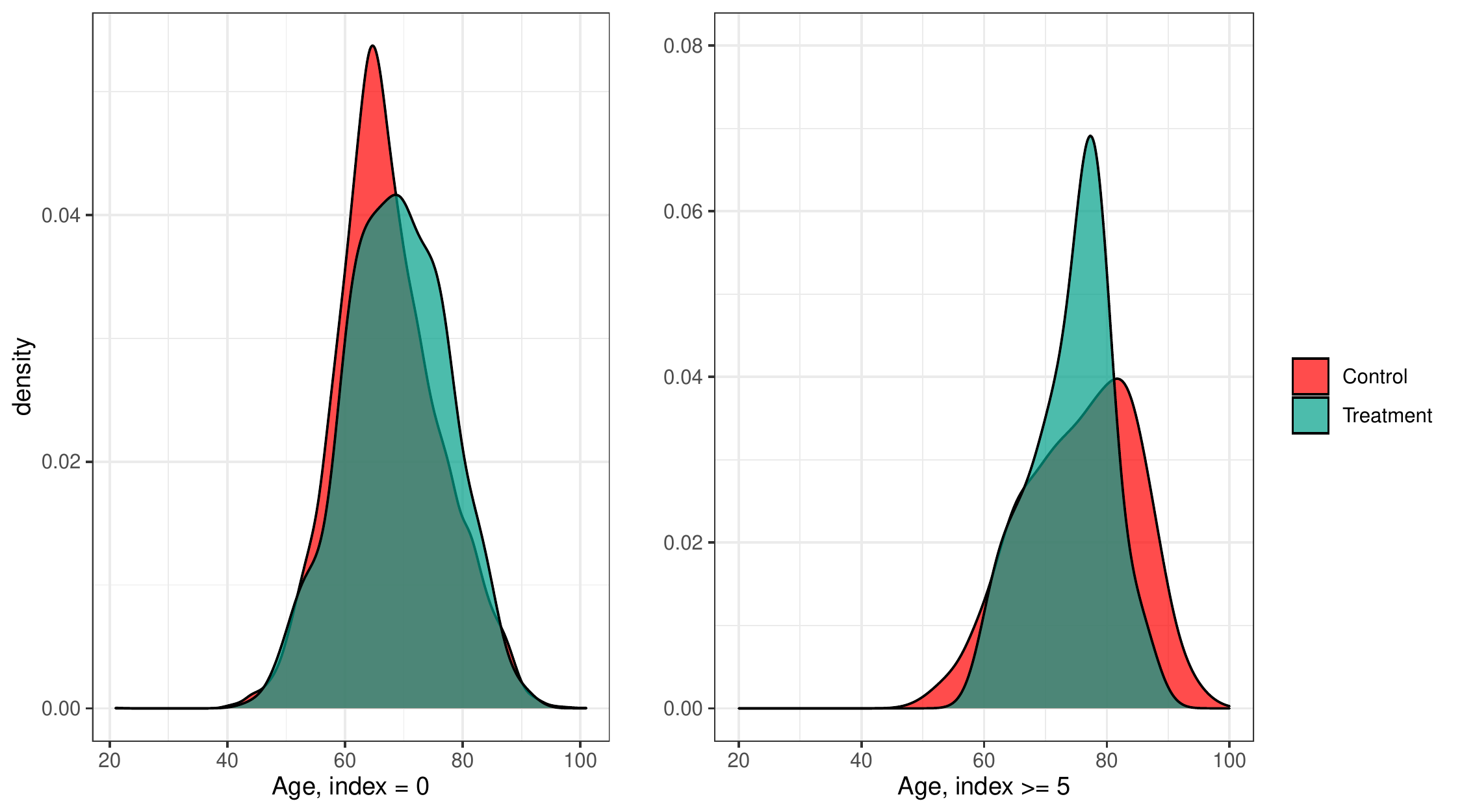} 
\caption{Age distribution when (1) Elixhouser index $= 0$ and (2) Elixhouser index $>= 5$}
\label{age}
\end{figure}

The fact that the NAM treated population is older than the comparison group may however distort the univariate analysis of these variables. To illustrate the problem, the distribution of age for the group with best health (left panel) and the group with worse health (right panel) is displayed in Figure \ref{age}. From the figure we can see that the patients with an Elixhouser index $\geq 5$ are older in the comparison group than in the NAM treatment group. This exercise shows the importance of flexible modeling, that is, to also allow for interactions among the covariates in the design.

\begin{table}[ht]
\caption{Means (standard deviations) of pre-diagnosis covariates of those prescribed NAM and comparisons (SoC) and the mean differences.}
\label{ttest}
\centering
\footnotesize
\begin{tabular}{lrrr}
Description & SoC & NAM & Difference \\
\hline
\hline
Age at diagnosis & 69.40 (9.16) & 70.85 (8.45) & -1.45$^{***}$  \\ 
Number of visits 1 month bf. diagnosis & 0.47 (0.84) & 0.71 (1.1) & -0.23$^{***}$  \\ 
Number of visits $1-6$ months bf. diagnosis & 1.38 (2.36) & 1.68 (3.15) & -0.30$^{***}$  \\ 
Number of visits $6-12$ months bf. diagnosis & 0.83 (1.96) & 0.82 (2.6) & 0.01 \\ 
Number of visits $1-60$ months bf. diagnosis & 7.12 (11) & 7.85 (23.72) & -0.73 \\ 
Elixhouser index 0, at diagnosis & 0.53 (0.5) & 0.44 (0.5) & 0.09$^{***}$  \\ 
Elixhouser index $1-4$, at diagnosis & 0.43 (0.5) & 0.51 (0.5) & -0.08$^{***}$  \\ 
Elixhouser index $5$, at diagnosis & 0.03 (0.18) & 0.05 (0.22) & -0.02$^{***}$  \\ 
Elixhouser index $0$, 12 months bf. diagnosis & 0.62 (0.49) & 0.54 (0.5) & 0.08$^{***}$  \\ 
Elixhouser index $1-4$, 12 months bf. diagnosis & 0.36 (0.48) & 0.43 (0.5) & -0.07$^{***}$  \\ 
Elixhouser index $5$, 12 months bf. diagnosis & 0.02 (0.13) & 0.03 (0.17) & -0.01$^{**}$  \\ 
Less than secondary school education & 0.38 (0.49) & 0.36 (0.48) & 0.02  \\ 
Secondary school education & 0.39 (0.49) & 0.39 (0.49) & -0.01  \\ 
Living with a partner & 0.65 (0.48) & 0.64 (0.48) & 0.01  \\ 
Born in the Nordic countries & 0.95 (0.22) & 0.94 (0.23) & 0.01  \\ 
\hline
\hline \\[-2ex]
\multicolumn{3}{l}{\textit{Standard deviations within parentheses.} $^{*}p<$0.1; $^{**}p<$0.05; $^{***}p<$0.01.} \\ 
& \\[-1.8ex] 
\end{tabular}%
\label{Elixhauser}
\end{table}

\subsubsection{Covariates measuring the health progression}
For each month after the diagnosis for the comparison patients and up until the month of treatment for those prescribed a NAM, we add the Elixhouser comorbidity index, the cumulative number of all hospital visits per month after diagnosis, the number of months with visceral metastases since diagnosis, the number of months with skeleton metastases since diagnosis and presence of metastases (0/1, taking value 1 if the patient is diagnosed with any metastases (ICD C77-C79) up until that month). 

How well patients responded to prior androgen deprivation therapy (ADT) is seen by experts as an important covariate. Therefore, we also include the total number of collected daily doses with a prescription with any of the drugs displayed in Table \ref{drugs3}. In addition, the indicator of being prescribed only Bicalutamide or Bicalutamide+GnRH up until a specific month is included.\footnote{The reason is that treatment with only bicalutamide can be given to patients with no symptoms. https://kunskapsbanken.cancercentrum.se/diagnoser/prostatacancer/vardprogram/primar-behandling-av-prostatacancer-med-spridning.}  Descriptive statistics of these variables will discussed in the section \ref{sec:Ident}.

\begin{table}[th]
\caption{Description of included ATC codes that are used in androgen deprivation therapy (ADT)treatment}\centering{\small 
\begin{tabular}{llr}
\hline
Group & Name & ATC cod(es) \\ \hline
Antiandrogen & Bikalutamid & L02BB03 \\ 
GnRH-agonist & Degarelix & L02BX02 \\ 
GnRH-analog & Buserelin & L02AE01 \\ 
& Leuprorelin & L02AE02 \\ 
& Goserelin & L02AE03 \\ 
& Triptorelin & L02AE04 \\ \hline
\end{tabular}
}
\label{drugs3}
\end{table}

\section{Study design and method}
\label{sec:Ident}

The design takes stock in the Neyman potential outcomes framework \citep{Neyman1923} and the Rubin Causal Model \citep{Rubin1974, holland1986} for
observational data. Define the potential outcome if a patient has been
given NAM $t$ months after being diagnosed for a PC $Y_{i}(1)$ and $Y_{i}(0)$
if he, instead, is given SoC. We observe outcomes $Y_{i},$ the treatment, $T_{i}$
and a large set of covariates, $\mathbf{Z}_{it}=(\mathbf{X}_{i},\mathbf{H}%
_{it})$ for a sample $n$ treated patients, where $\mathbf{X}_{i}$ are the
covariates measured before the diagnosis and $\mathbf{H}_{it}$ are the health
progression covariates. 

Our interest is that of estimating the sample average treatment effect of
the treated, formally defined as
\begin{equation}
ATET=\frac{1}{n}\sum_{i=1}^{n}\{Y_{i}(1)-Y_{i}(0)|T_{i}=1\}. \label{ATE}
\end{equation}%

Note that we here neglect that the treatment effect could differ across the
month of treatment after being diagnosed for a PC. However, heterogeneity in effect across $t$ will be investigated in the analysis (see Section \ref{sec:Plan}). Note that, as we observe the outcome for at least 24 months, the ATET can be estimated for each month after prescription for up to two years. 

In order to consistently estimate the ATET, $Y_{i}(1)$ and $Y_{i}(0)$\ are
not allowed to be affected by the treatment given (known as the
stable unit treatment value assumption, SUTVA). The patients are being treated in different hospitals and at many different times why spillover effects between patients should be low, thus we assume SUTVA is valid.\footnote{As this is an observational study, the validity of the SUTVA is most likely higher than in a non-blinded randomized control trial (RCT). Being part in an experiment affect peoples behavior. This is the reason for blinding and hence to give a placebo. Thus, SUTVA may not be a realistic assumption in many non-blinded RCT \citep{Teira_2013}. Interestingly, the assumption may not also be valid in blinded RCT:s. \citep{Teira_2013,Schuklenk_2003} discuss the spillover effects in blinded RCT:s analyzing effects of drugs to patients suffering from AIDS.}  The implication is that $Y_{i}(1)$ and $Y_{i}(0)$
are the observed outcomes for those prescribed a NAM and the SoC, respectively.

The fundamental problem estimating the ATET defined in \eqref{ATE} is that
we do not observe $Y_{i}(0)$ for those prescribed NAM. Under the weak
unconfoundedness assumption (UA): 

\begin{equation}
\Pr (T_{i}=1|\mathbf{Z}_{it},Y_{i}(0))=\Pr (T_{i}=1|\mathbf{Z}_{it})=p(%
\mathbf{Z}_{it}), \label{PS}
\end{equation}%
where $p(\mathbf{Z}_{it})$ is known as the propensity score, the
counterfactual means for those prescribed a NAM can be calculated given
that we can find comparison patients to the NAM patients with the same $%
\mathbf{Z}_{it}$ (known as the common support assumption). The UA imply
that $\mathbf{Z}_{it}$ should contain everything that jointly determine $%
T_{i}$ and $Y_{i}(0)$. Hence, any difference in observed outcomes between the
NAM treated and comparions if not being prescribed a NAM must be caused by the
treatment holding $\mathbf{Z}_{it}$ constant. The implication of these three
assumptions is that an individual $i$ prescribed a NAM in month $t$ with
covariates $\mathbf{Z}_{it}$ is not systematically different from patient 
$j$ in the comparison group if $\mathbf{Z}_{it}=\mathbf{Z}_{jt}$. That is,
the reason why patient $i$, but not patient $j$, is prescribed the
NAM is random, like in a randomized experiment. The main argument for the
randomness here is that the comparison patients where not offered the drugs at the time when they had the PC diagnosis.

The NAM treated are patients suffering from mCRPC. This means they constitute a highly selective group with severe health problems. As a consequence, it is important to validate if there are patients in the comparison group with similar health progression as in the treatment group. If not, the common support assumption is not valid.

\begin{figure}[tbp]
\centerline{
\makebox[\textwidth][c]{\includegraphics[width=.8\textwidth]{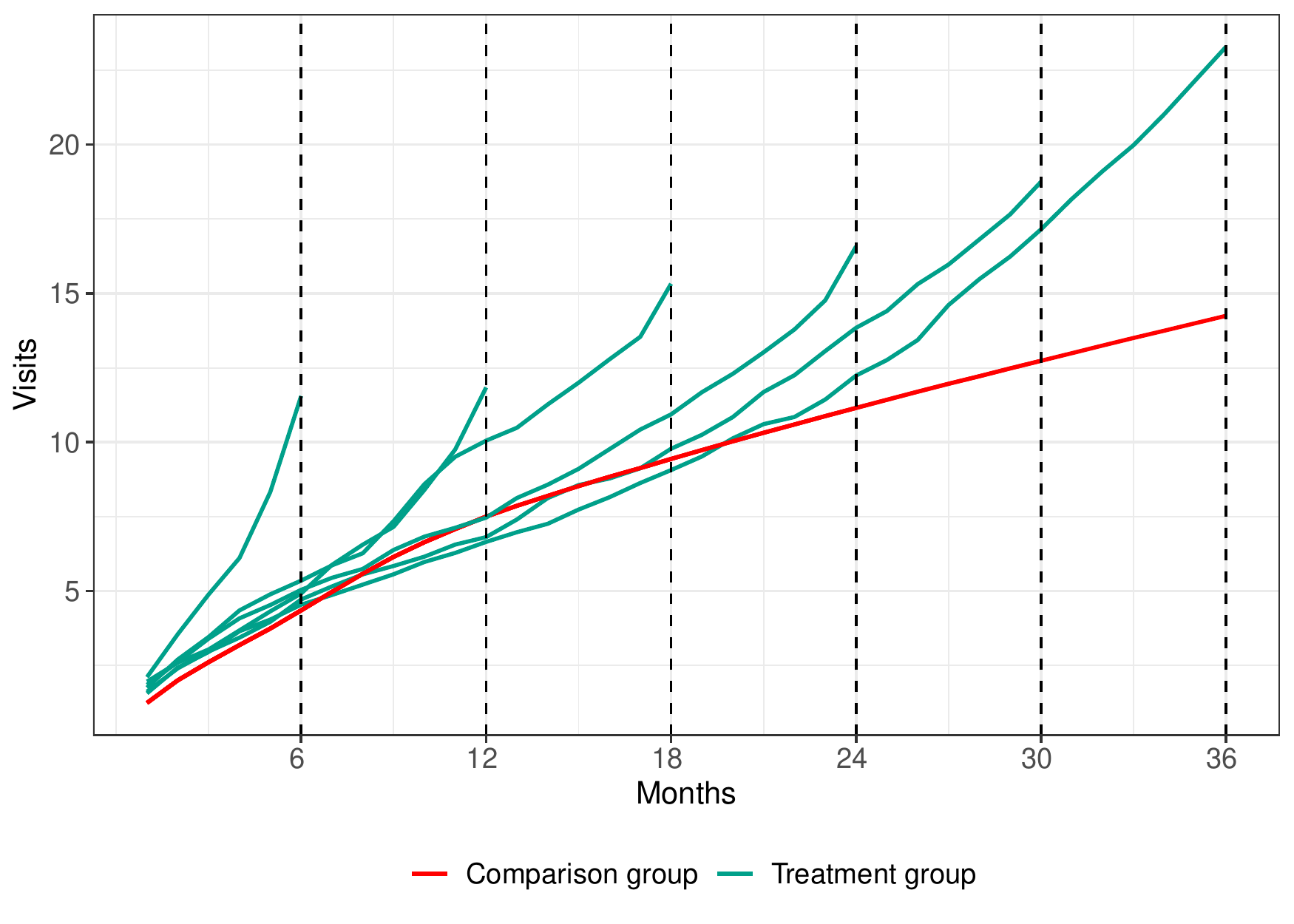}} 
}
\caption{Average number of visits for those NAM treated 6, 12, 18, 24, 30 and 36 months after diagnosis and the comparison group.}
\label{VISITS}
\end{figure}

Figure \ref{VISITS} shows the mean number of hospital visits for the NAM treated (the number of NAM treated is given within parenthesis) at month 6 (9), 12 (25), 18 (37), 24 (47), 30 (32) and 36 (47) after the diagnosis, together with the mean number at the same periods of
the $19,456$ patients in the comparison group. Each line corresponds to the
progression for each of the seven groups. The figure shows the importance of
considering the health in the DTP and confirms that the health progression for the NAM treated group is
less good than for the comparison group. The figure also reveals that
patients prescribed NAM early have a larger increase in the number of visits
compared to those prescribed NAM later. This provides suggestive evidence that those
prescribed early has a less good health progression than those prescribed
later. There is substantial overlap in the
distributions of hospital visits for all months. The distribution for the months displayed
in Figure \ref{VISITS} is displayed in the Appendix (Figure \ref{VISITS2}). This means that it, most likely, will be possible to find men in the comparison group with the same health progression as the NAM treated.

\begin{figure}[]
\centerline{
\makebox[\textwidth][c]{\includegraphics[width=0.8\textwidth]{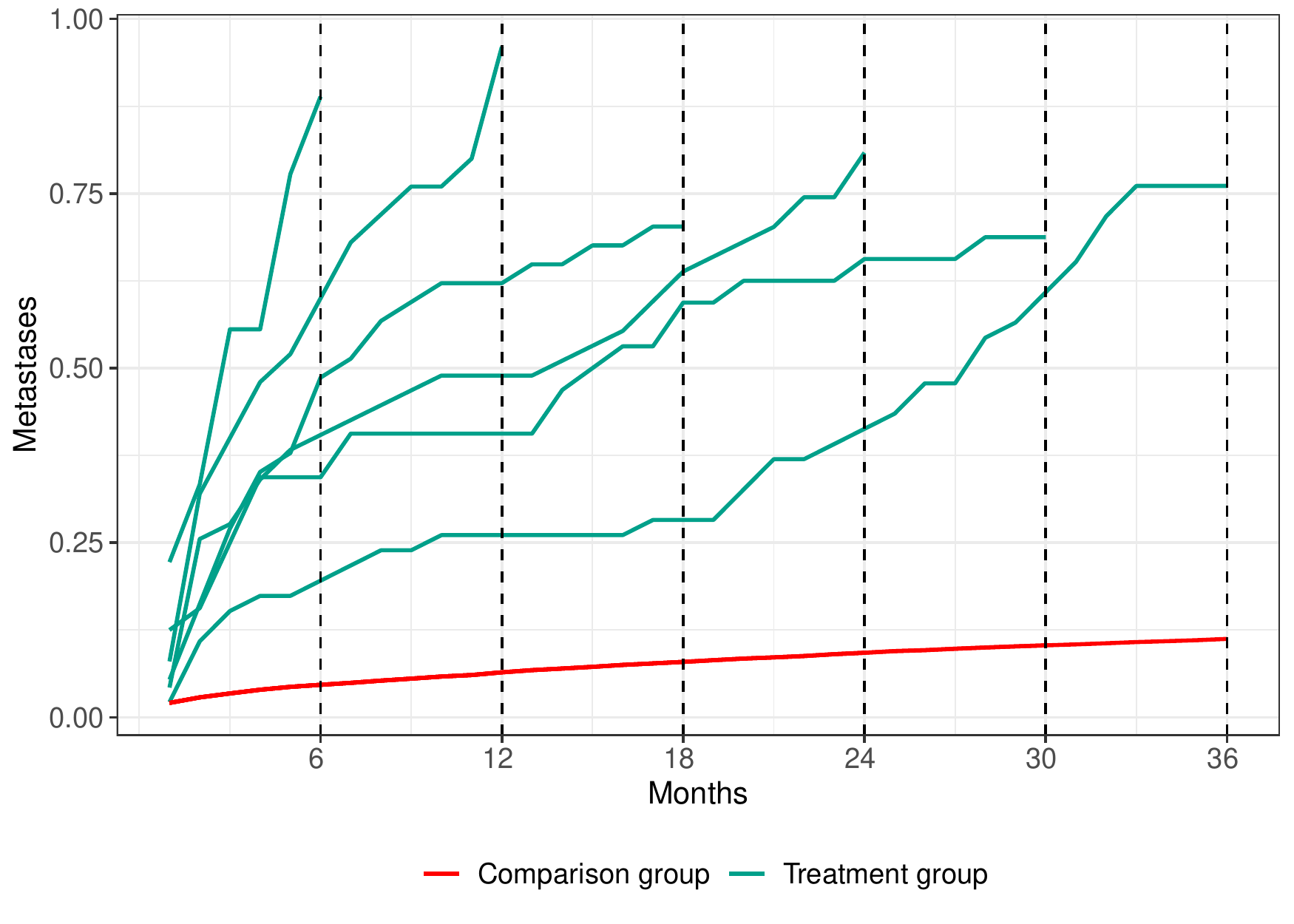}} 
}
\caption{Fraction of patients with metastases for those NAM treated 6, 12, 18, 24, 30 and 36 months after diagnosis and the comparison group.}
\label{METAS}
\end{figure}

In total, almost 90 percent of the NAM treated have had visits related to metastases before being prescribed a NAM (in total 30, 9 and 82 percent have inpatient visits related to ICD codes C77, C78 and C79 respectively). The cumulative fraction with a C77, C78 or C79 diagnose each month after the diagnose for NAM treated 6, 12, 18, 24, 30 and 36 months after diagnosis and for the comparison group is presented in Figure \ref{METAS}. 

We can see the same pattern as with hospital visits. Patients prescribed NAM early are also diagnosed with a metastasis early. While almost 90 percent of the nine NAM treated who was prescribed a NAM six months after diagnosis have metastases the month preceding the treatment about 5 percent in the comparison group have metastases at the same period. This means that close to 900 of the patients in the comparison group are suffering from metastases at the same period. Thus, we have around 10 possible comparison patients to each NAM treated six month after diagnosis. After 35 months, 11 percent in the comparison group have had one of the metastasis diagnoses. 

\begin{figure}[]
\centerline{
\makebox[\textwidth][c]{\includegraphics[width=.8\textwidth]{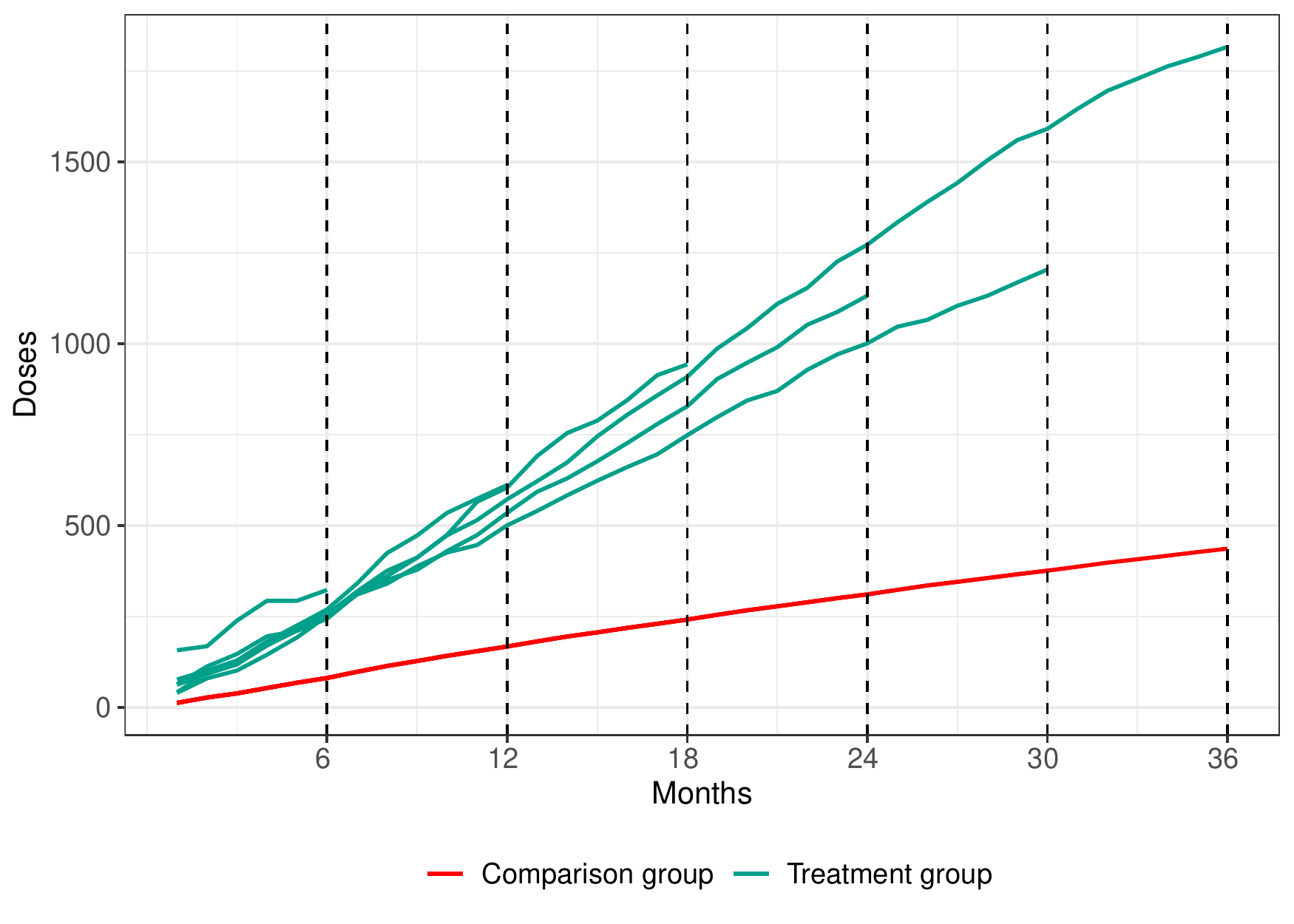}} 
}
\caption{Average number of daily doses for those NAM treated 6, 12, 18, 24, 30 and 36 months after diagnosis and the comparison group.}
\label{MEDS}
\end{figure}

Figure \ref{MEDS} shows the cumulative number of collected daily doses of ADT (see Table \ref{drugs3}) for those NAM treated 6, 12, 18, 24, 30, 36 months after diagnosis and for comparison. Those NAM treated after 36 months have on average collected more than $2\,000$ daily doses, compared to the average of 460 daily doses in the comparison group after 36 months\footnote{Note that the number of daily doses in many cases are greater than the number of days. This is, for example, due to the fact that the defined daily dose (ddd) of Bicalutamide is 50 mg, but prostate cancer patients can be prescribed 150 mg/day}. There is, as with hospital visits, substantial overlap in the distributions
for all strata. The distribution for the months displayed in Figure \ref{MEDS} is displayed in the
Appendix (Figure \ref{DAYS2}).

\begin{figure}[]
\centerline{
\makebox[\textwidth][c]{\includegraphics[width=.8\textwidth]{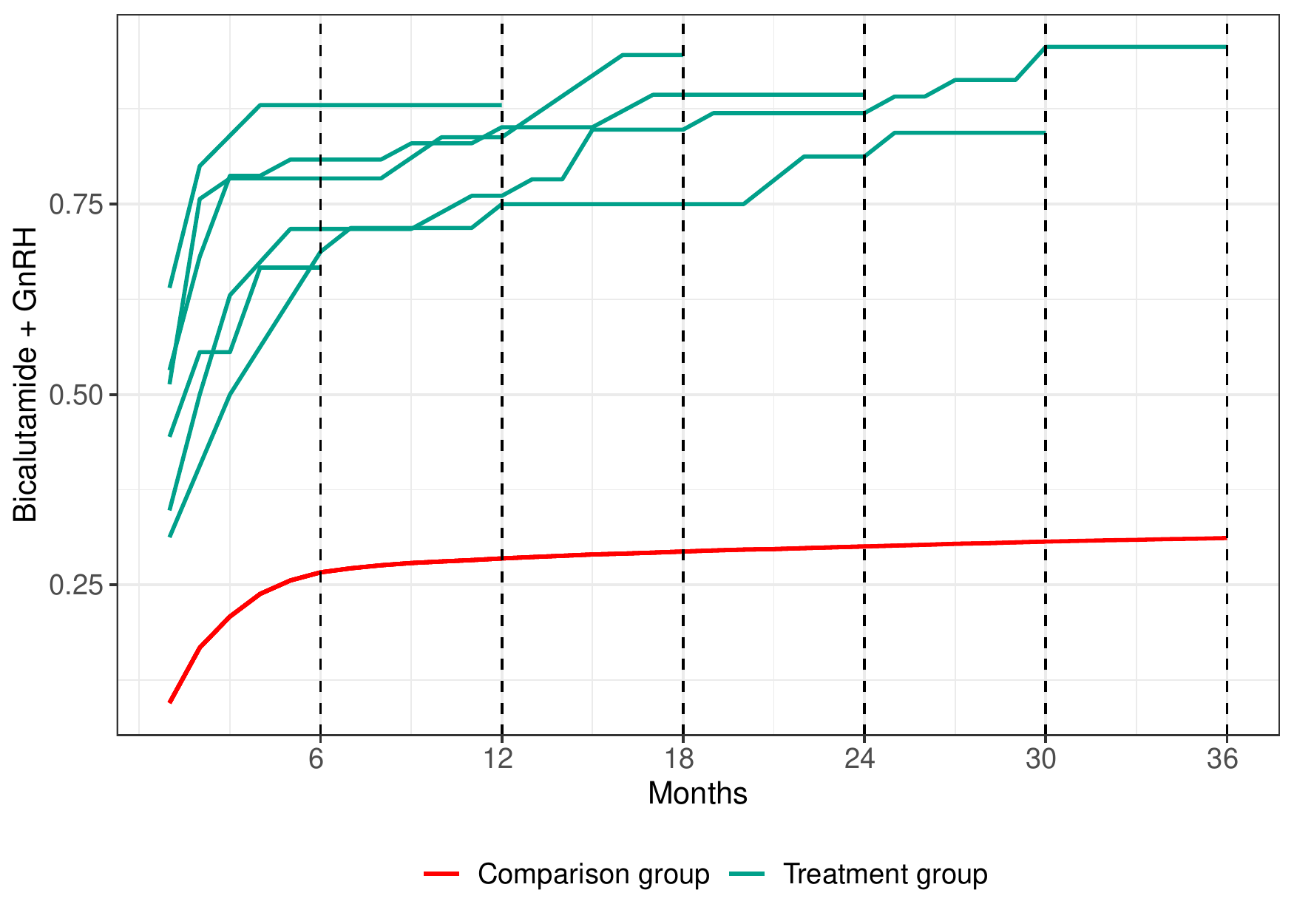}} 
}
\caption{Fraction of patients prescribed both bicalutamide and a GnRH for those NAM treated 6, 12, 18, 24, 30 and 36 months after diagnosis and the comparison group.}
\label{BIKA}
\end{figure}

As the different drugs can be administered differently depending on how severe the disease is, we do in addition look at the fraction who got both bicalutamide and a GnRH up until 36 months after diagnosis. The result is presented in Figure \ref{BIKA}. The pattern is the
same as above and given the large number of comparison patients and the small number of
NAM treated in each of the strata, there is, unconditionally, no problem finding patients
in the comparison groups with similar health progression. 

\subsection{Entropy balancing }

As we have a very large set of covariates, of which many are continuous, it
is not possible to match exactly on the covariates. An often used strategy
is to estimate the probability to be treated, that is, the (estimated)
propensity score (cf. equation (\ref{PS})), at each month and to find
matches based on this scalar. This means that with respect to this estimated
propensity score, the most similar comparison individual to a treated
individual in month $t$ will form a pair. This is known as a one-to-one
propensity score matching without replacement strategy in the literature.
The propensity score matching strategy can be time consuming as the final
specification of the regression model is determined iteratively until we
have obtained balance in the covariates between the two groups. In this
paper we instead make use of an entropy balancing scheme. Entropy
balancing tackles the adjustment problem from the reverse and estimates
weights (can be seen as the propensity scores) directly from imposed balance
constraints on the covariates. In contrast to the one-to-one propensity
matching without replacement, this means that some comparison individuals
are used multiple times. 

As the balancing is done for all $w_i$, $i=4...,36$, different sets of covariates are used for each $w_i$. We include, in addition to the chosen subset of covariates, interactions and polynomials. We also require balance with regard to the variances of the main covariates. In a few cases, the tolerance level in the entropy balance algorithm is not reached. As the standardized mean difference for the balanced samples is well below 0.25 (the rule of thumb suggested in \cite{Imbens_Wooldridge_2009}) for the all covariates, balance on all covariates can still be assumed.

\begin{figure}
\centering
\begin{subfigure}[b]{1\textwidth}
\centering		\includegraphics[scale=0.6]{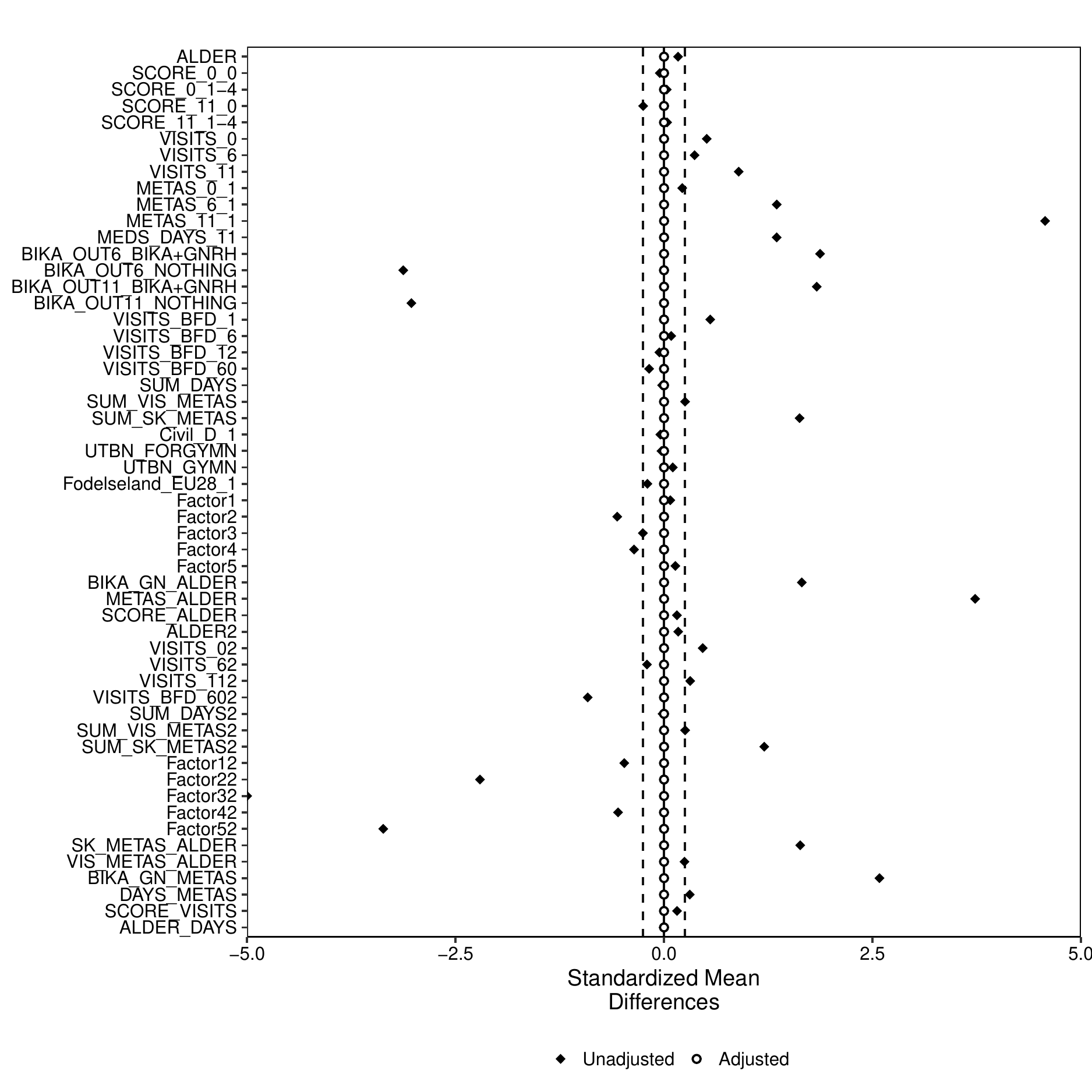}\\	        \caption{Standardized mean difference, 12 months}
		\label{Bal1}
\end{subfigure}
     \hfill
\begin{subfigure}[b]{1\textwidth}
 \centering	\includegraphics[scale=0.6]{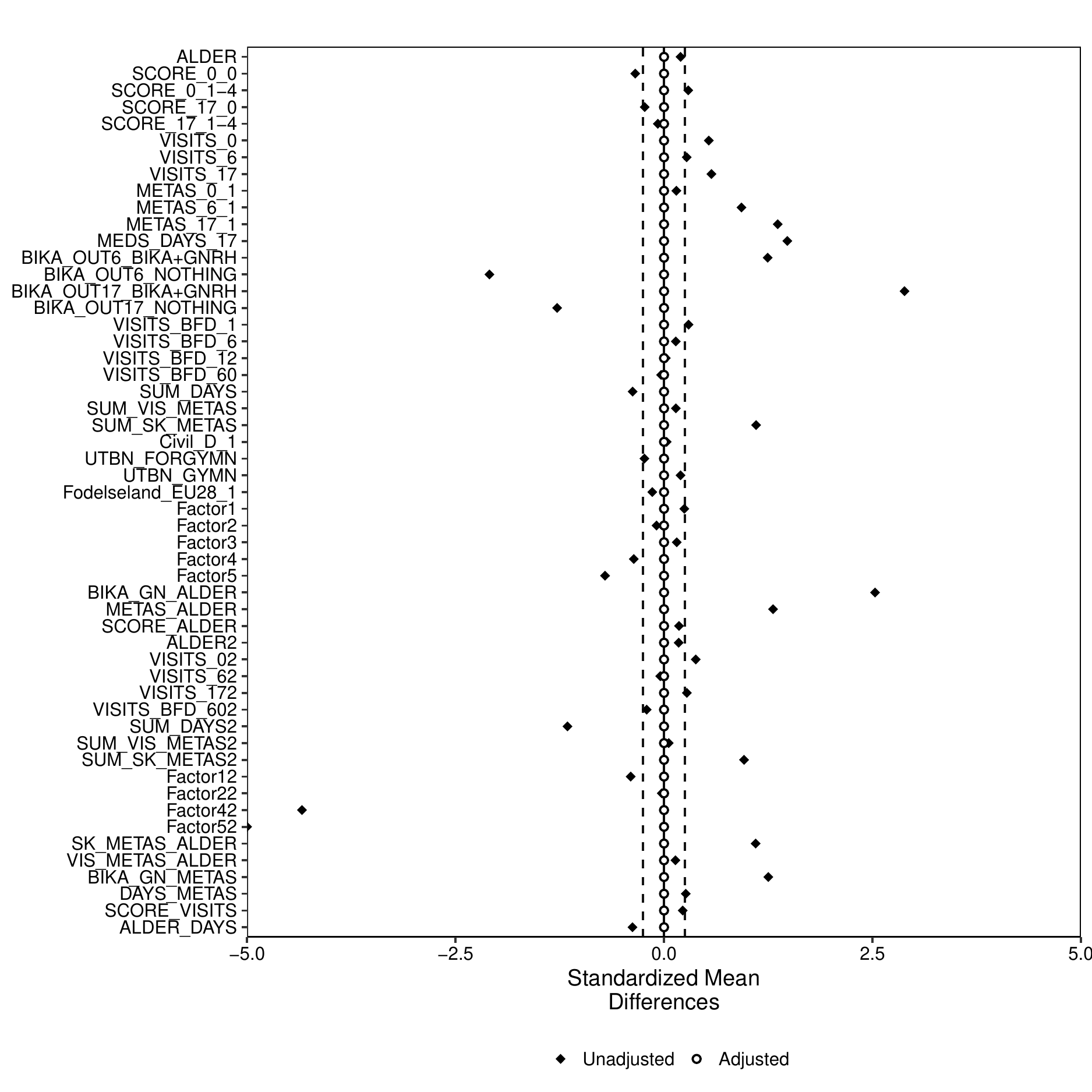}		        \caption{Standardized mean difference, 18 months}
\label{Bal2}
\end{subfigure}
        \caption{Balance, 12 and 18 months to treatment}
        \label{Bal22}
\end{figure}

\begin{figure}
\centering
\begin{subfigure}[b]{1\textwidth}
\centering
	\includegraphics[scale=0.6]{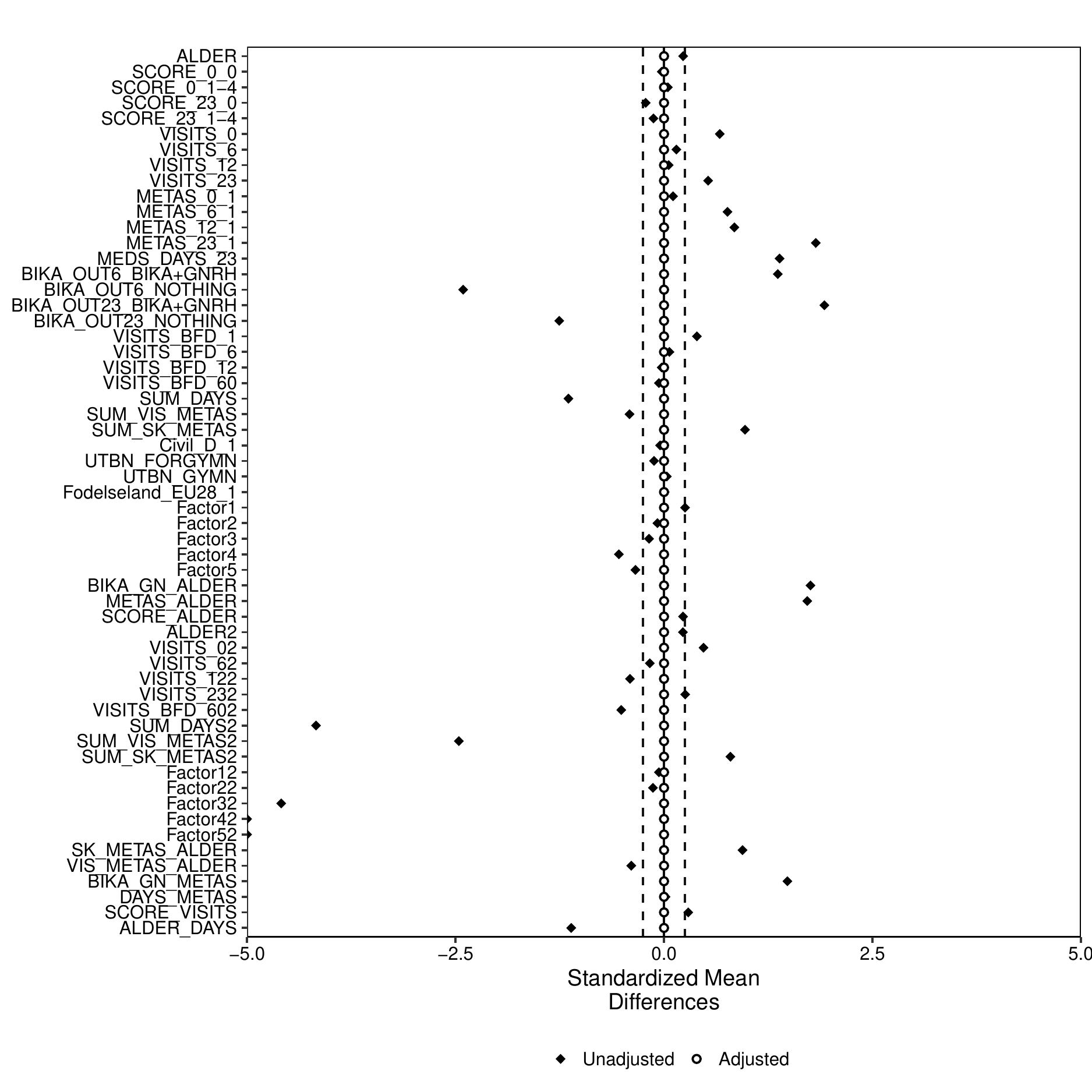}\\
		        \caption{Standardized mean difference, 24 months}
		\label{Bal3}
\end{subfigure}
     \hfill
\begin{subfigure}[b]{1\textwidth}
 \centering
		\includegraphics[scale=0.6]{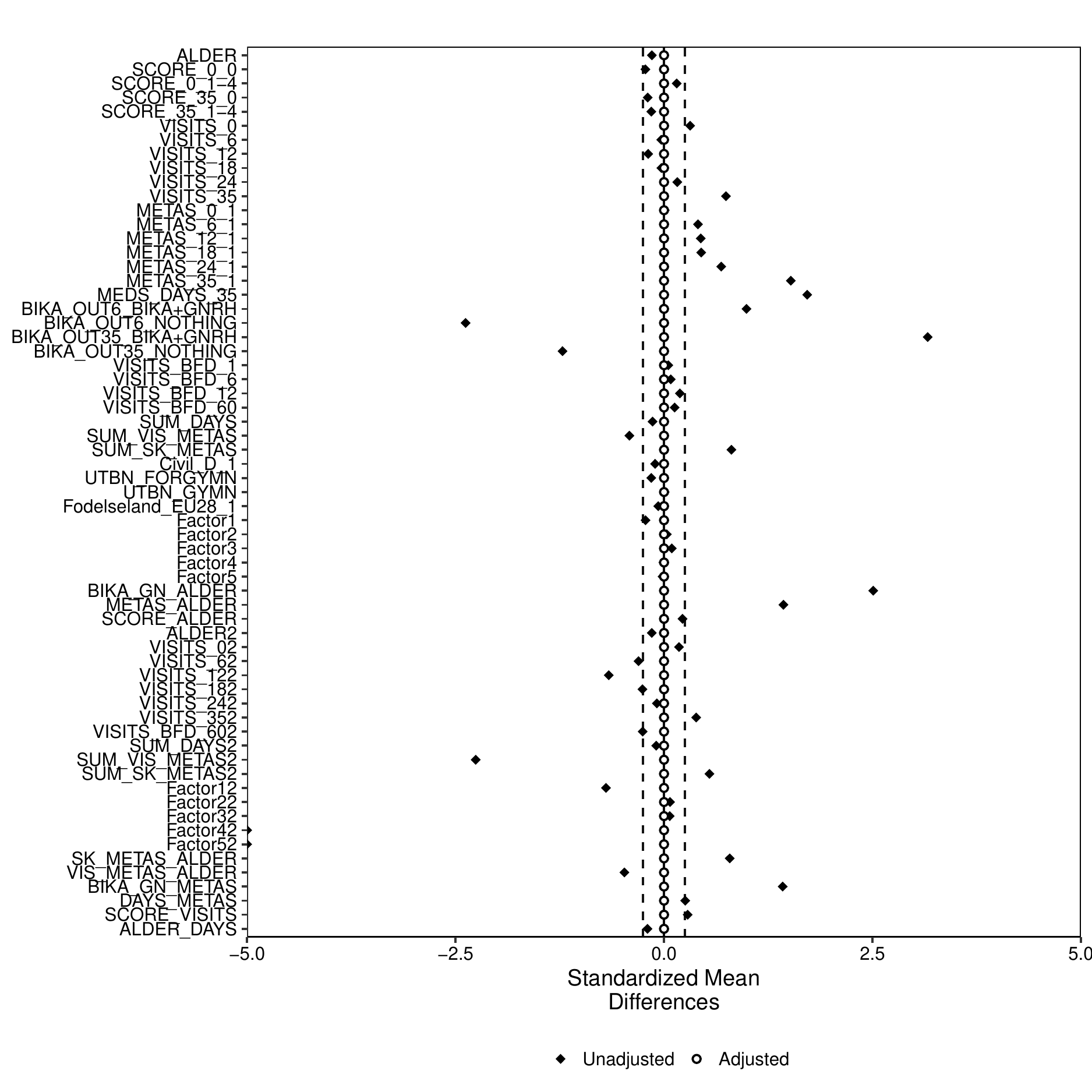}
		        \caption{Standardized mean difference, 36 months}
\label{Bal4}
\end{subfigure}
        \caption{Balance, 24 and 36 months to treatment}
\end{figure}

Examples of for the subgroups with a time until treatment of 12, 18, 24 and 36 months are presented in Figure \ref{Bal1}, \ref{Bal2}, \ref{Bal3} and \ref{Bal4}. We can see that there are mean differences between the groups, but these differences are, as expected, removed when comparing the treated when weighting the comparisons, i.e. the implicit control group. 

\begin{figure}
\centering
\begin{subfigure}[b]{0.4\textwidth}
\centering
		\includegraphics[scale=0.6]{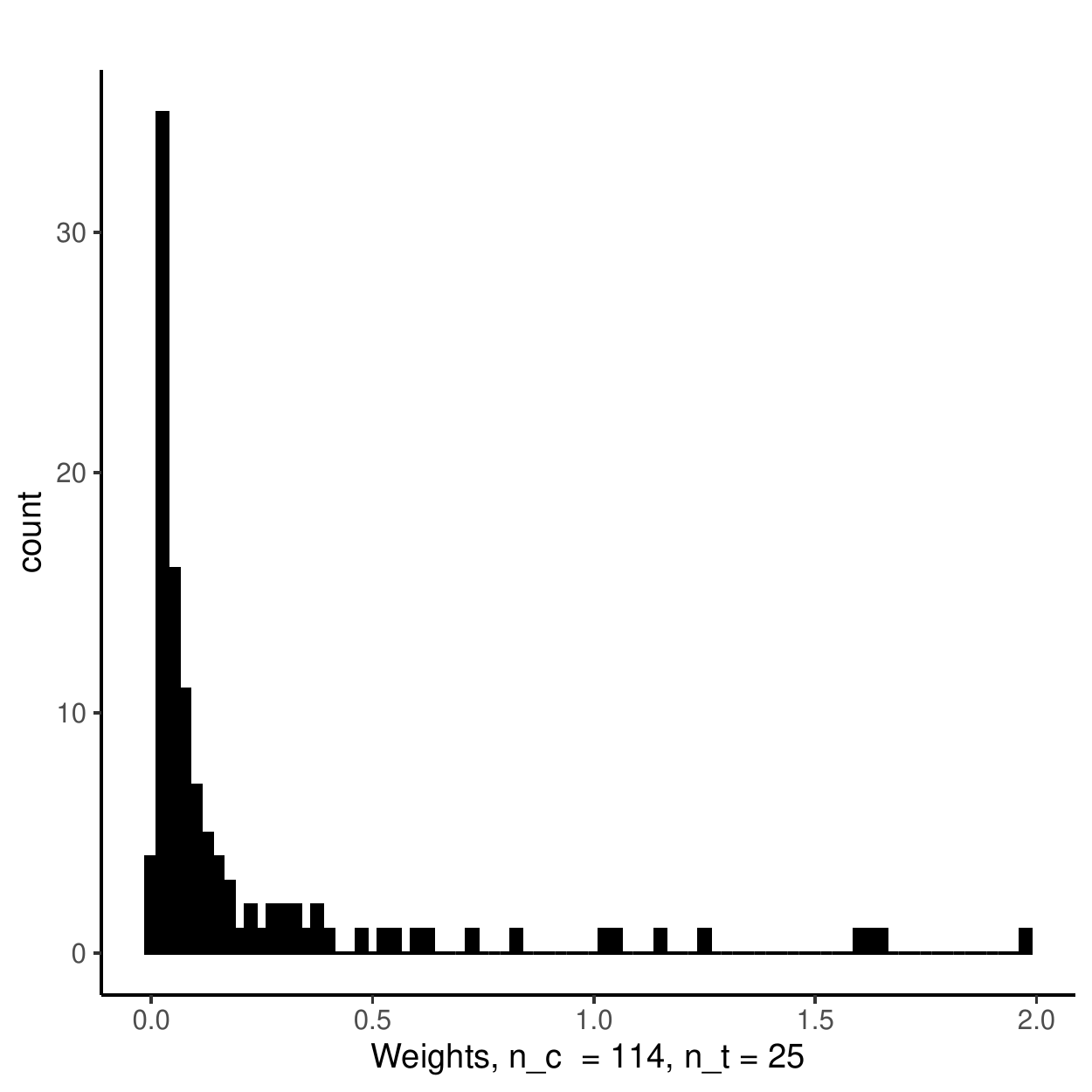}\\
		        \caption{Weigths, 12 months}
		\label{W1}
\end{subfigure}
     \hfill
\begin{subfigure}[b]{0.4\textwidth}
 \centering
		\includegraphics[scale=0.6]{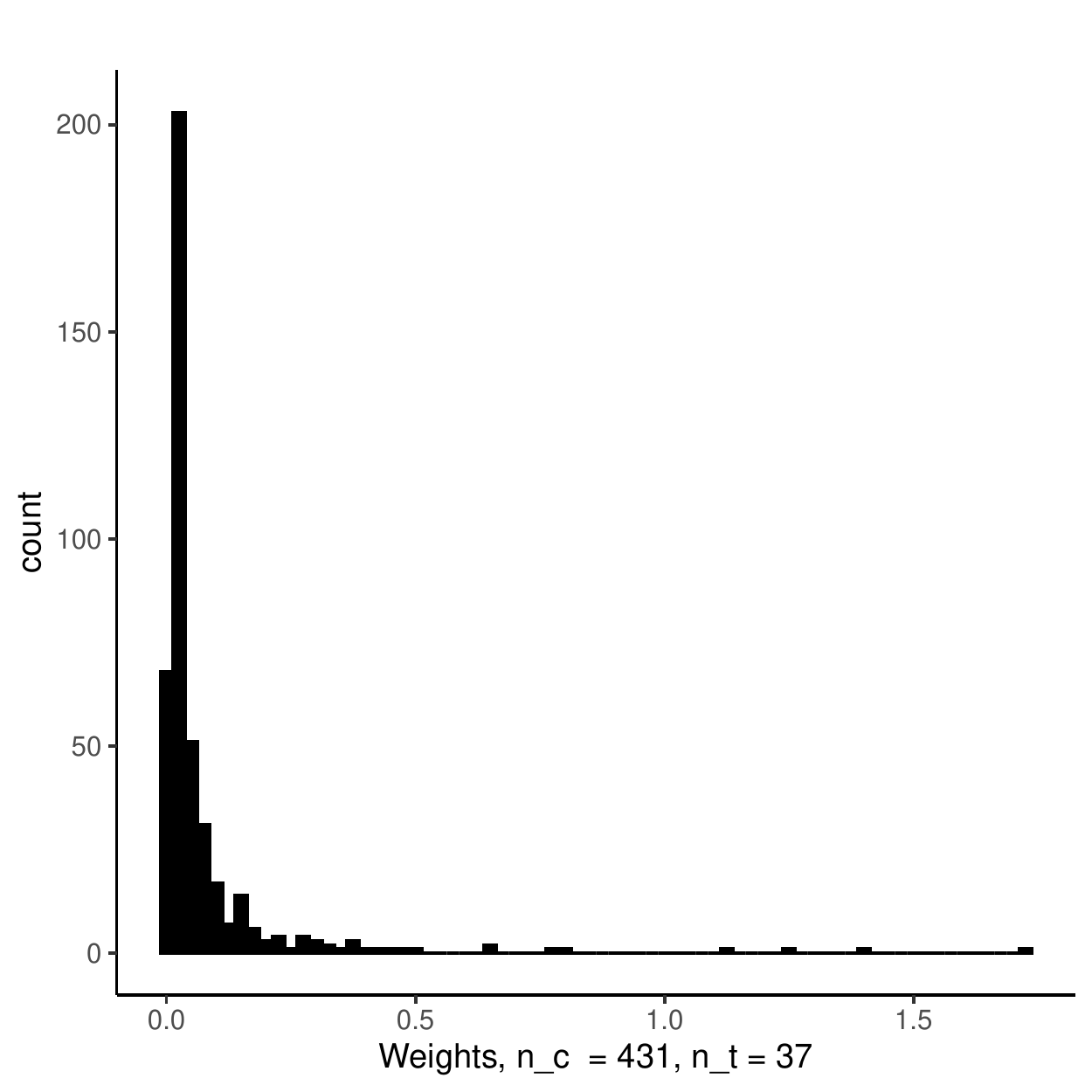}
		        \caption{Weigths, 18 months}
\label{W2}
\end{subfigure}
        \caption{Weights, 12 and 18 months to treatment. The figure shows the distribution of
the weights for the 114 and 431 observations in the control group with a weight greater
than 0.01. All, in this case, 25 and 37 NAM treated individuals respectively have weights
equal to one after balancing and are not shown in the figure.}
\end{figure}

After re-weighting, the total weight of the comparison group exactly matches that of the treatment group. The distribution of the estimated weights is skewed and only a small fraction, about 1-2 percent, got an estimated weight above 0.01. The weights for this fraction is presented in Figure \ref{W1} and \ref{W2}. As we knew that the groups differ substantially, the many zero-weights are expected. It is thus of importance that a single control observation does not get an extremely high weight.

\begin{figure}
\centering
\begin{subfigure}[b]{.4\textwidth}
\centering
		\includegraphics[scale=0.6]{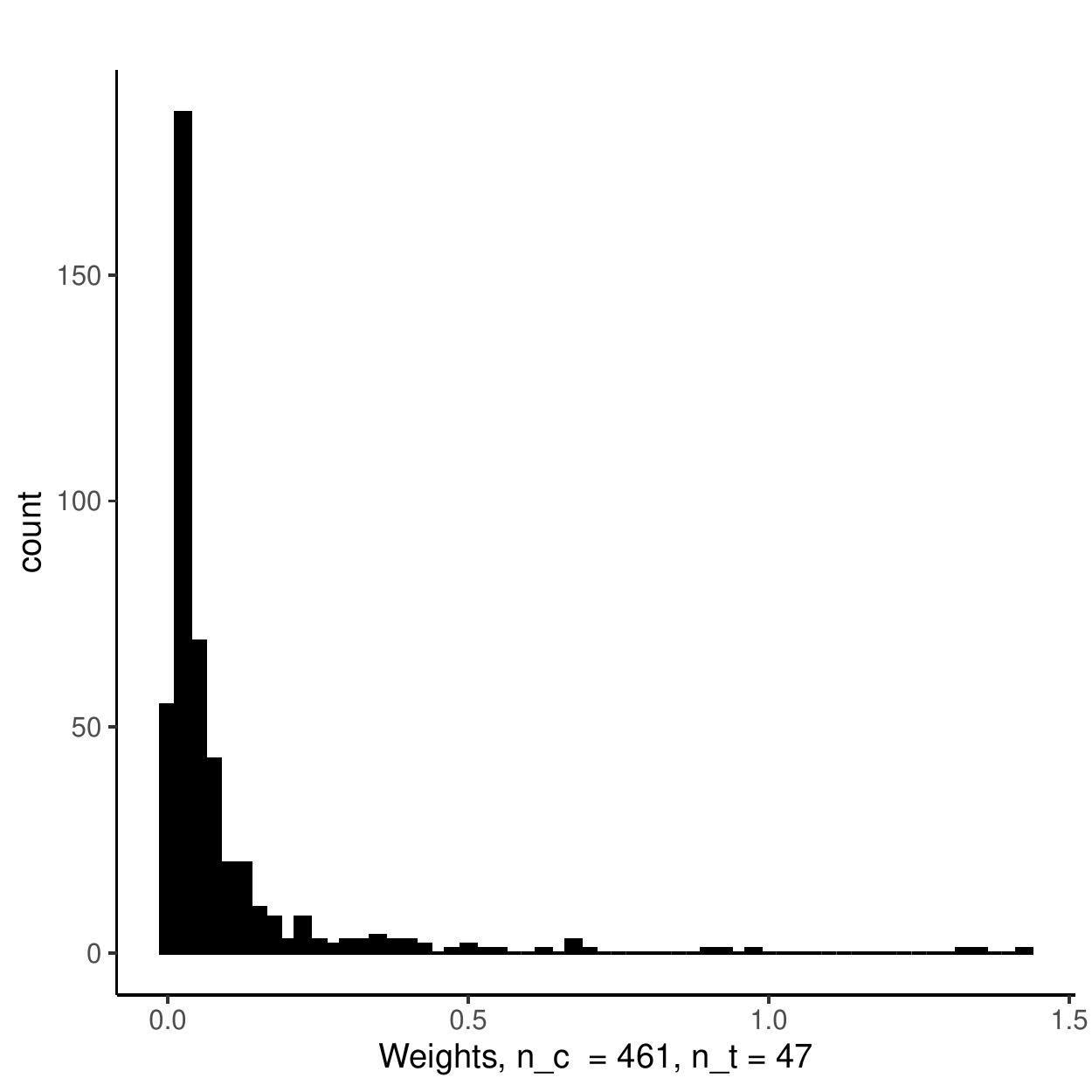}\\
		        \caption{Weigths, 24 months}
		\label{W3}
\end{subfigure}
     \hfill
\begin{subfigure}[b]{.4\textwidth}
 \centering
		\includegraphics[scale=0.6]{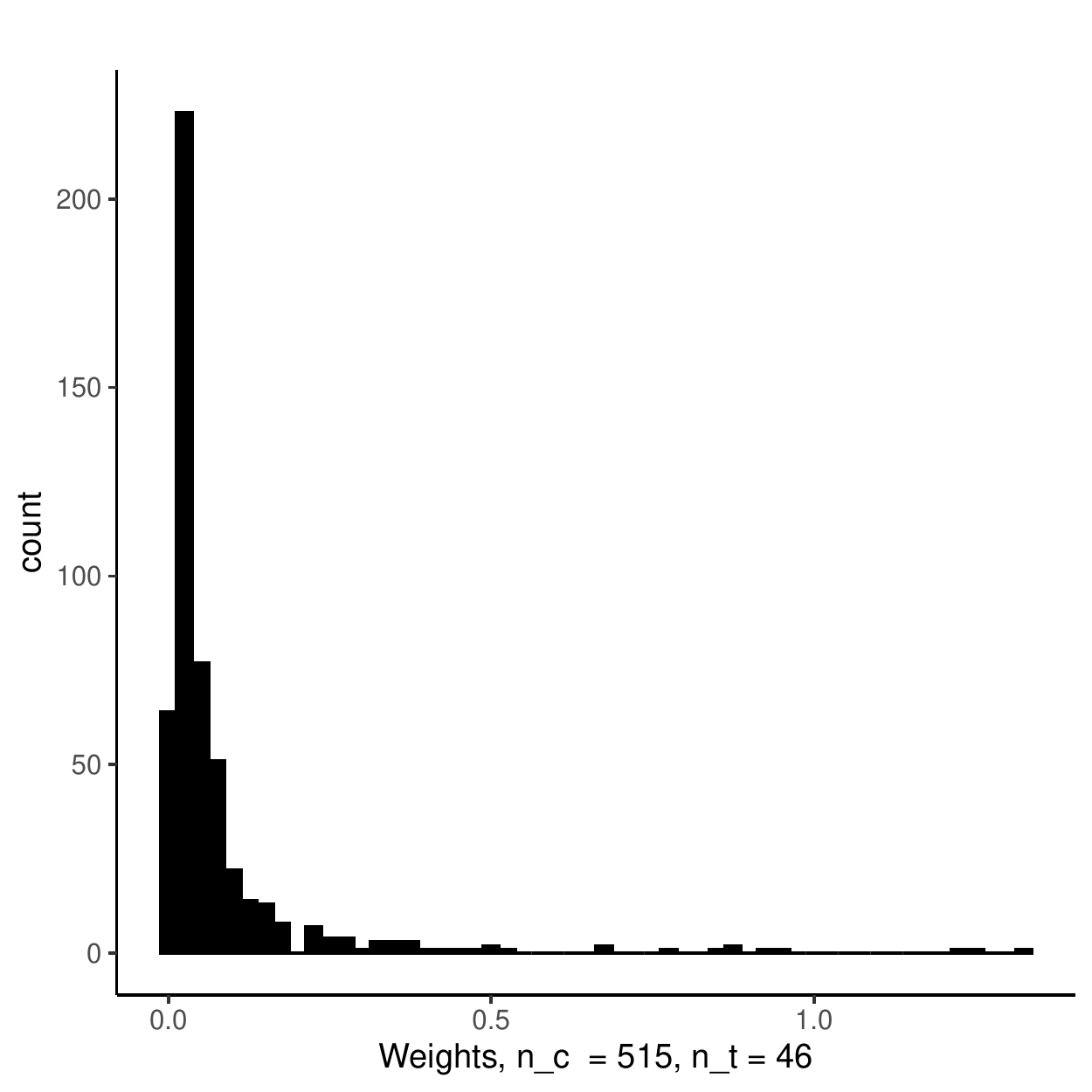}
		        \caption{Weigths, 36 months}
\label{W4}
\end{subfigure}
        \caption{Weights, 24 and 36 months to treatment. The figure shows the distribution of
the weights for the 461 and 515 observations in the control group with a weight greater
than 0.01. All, in this case, 47 and 46 NAM treated individuals respectively have weights
equal to one after balancing and are not shown in the figure.}
\end{figure}

\begin{figure}[tbp]
\centerline{
\makebox[\textwidth][c]{\includegraphics[width=.75\textwidth]{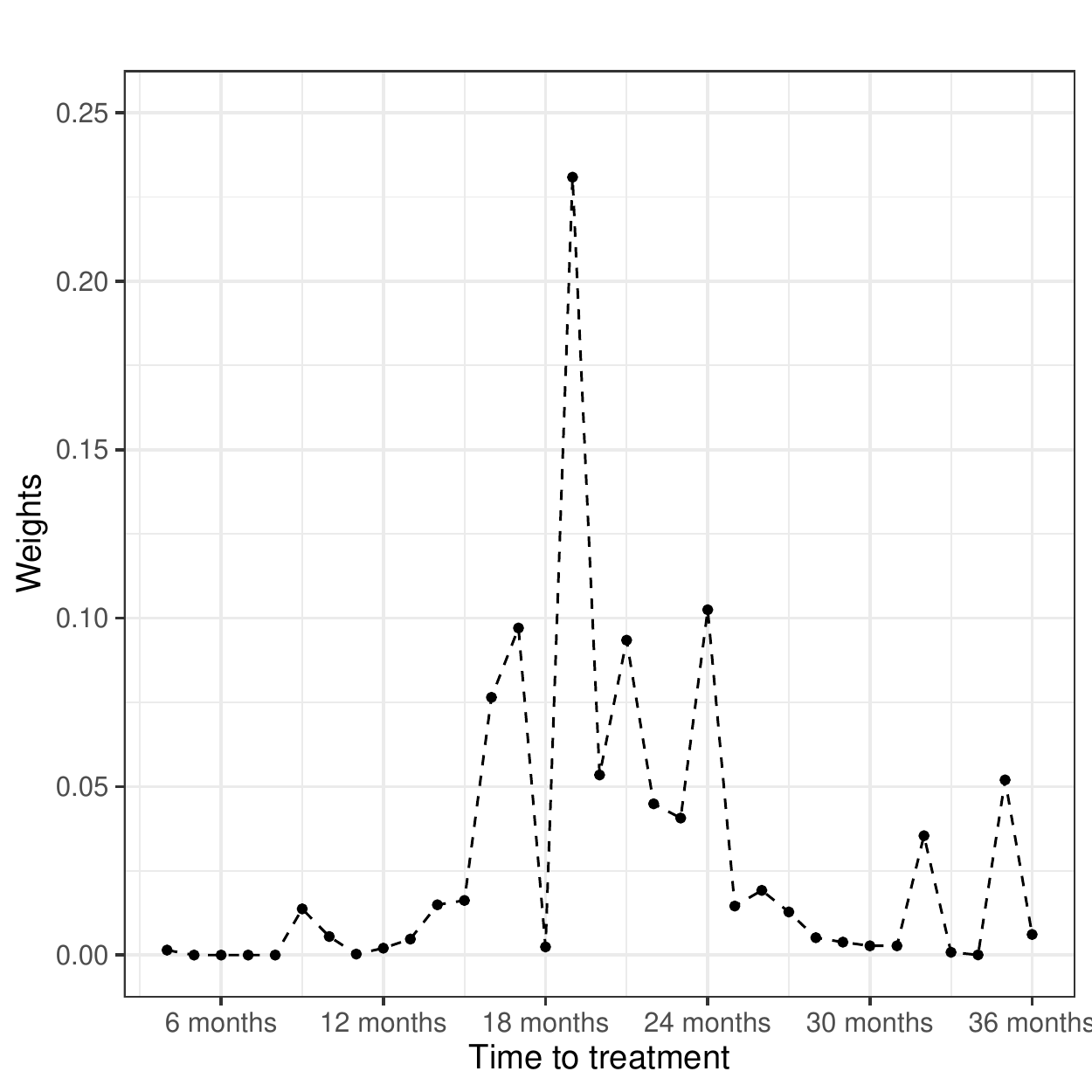}} 
}
\caption{The different weights for $w_j$, $j=4,...,36$, for the comparison patient described in Figure \ref{Fig:Control_ind}.}
\label{WW}
\end{figure}

Every comparison individual gets in other words a weight conditioned on each of the $t=4,...,36$ time points from diagnosis to treatment. Turning to the man with the health progression displayed in Figure \ref{Fig:Control_ind}, who was 78 years old when diagnosed with PC. He was diagnosed with skeleton metastases within 6 months after the diagnose and had heart related problems and an Elixhouser score of 3. Furthermore, he was treated with both bicalutamide and a GnHR (indicating a fast progression). The weights over the three years given to this
patient is displayed in Figure \ref{WW}. The largest weight is given between 12 and 24 months, which thus seems reasonable as the disease progression
for this man seems to be quite fast.

\section{Planned analysis}
\label{sec:Plan}

As we did not have access to mortality data in the design of the study
protocol, there is a risk that a comparison individual is dead at the time
when being assumed to be given the SoC treatment. This is not a
problem for the NAM treated as we know that he is alive when being prescribed
the NAM. For this reason we will after we have added the mortality data
remove comparisons from the analysis who is found to have died before the
imputed time of the SoC treatment. 

We have one primary, and two secondary outcomes. The primary outcome is
all-cause mortality (DEAD) and the two secondary outcomes capturing
morbidity are PAIN, and SRE. PAIN is an indicator for severe pain, and SRE
is an indicator for a skeleton related events. The reason for including PAIN and SRE as outcomes is that these two
morbidity outcomes have been seen as common complications of bone metastases.

DEAD as a primary endpoint is an indicator variable defined as one for
patients who are dead of any causes and zero for other patients at the end
of each 30 days period after being prescribed a NAM. Mortality data will be
available until the end of June, 2020. 

Patients are assumed to suffer severe pain if they receive prescriptions for
opiates in combination with tramadol and paracetamol
(ATC-codes N02AA, N02AX02, and N02BE01). The PAIN indicator is one for
periods in which the patient has received such a subscription and zero for
the other periods. Prescription data will be available until 31st of
December, 2020. 

Patients are assumed to suffer skeleton related event if they experience a
hospitalization because of pathologic fracture (ATC codes M485, M495, M844,
and M907) or spinal cord compression (G550, G834, G952, G958, G959, and
G992) (Parry et al., 2019). The SRE indicator is one for periods with such
hospitalizations and zero for the other periods. As inpatient care data is
available until 31st of December, 2019. 

For all outcomes a weighted least squares (WLS) regression analysis will be used
in the estimation of ATET. Since entropy balancing orthogonalizes the
treatment indicator with respect to the covariates that are included in the
balance constraints we do not add any covariates but
we control for the treatment month, $t$. This means that we are \ estimating 
\begin{equation}
Y_{im}=\alpha _{m}+\alpha _{t}+\beta _{m}T_{i}+\varepsilon _{im},m=1,...,24%
\text{ (36)} \label{OLS}
\end{equation}%
using WLS, with the weights for each observations obtained from the entropy balancing
algorithm. Here $m$ is the index for the month of outcome, $\alpha _{t}$ are
the parameters controlling for the month of treatment and $\beta _{m}$ is
the ATET effect at month $m$. \ For the inference we will use the
Eicker-Huber-White (EHW) covariance estimator \citep{Eicker_1967,Huber_1967, White_1980}  robust covariance matrix. 

For the inference we will use Bonferroni correction with a five percent
overall level. With one primary and two secondary outcomes this means that
the significance level on the single outcomes will be 1.67\% (= 100*0.05/3).
For MORTALITY and PAIN we can evaluate the effect on all patients for up to
24 months. 

The problem with analysis of the two morbidity outcomes is that they are
only observed in data if the patient is alive. In the analysis above we
will, at each evaluation period, remove the dead patients from the analysis.
This means that the number of valid observations will be reduced over the
evaluation window. If there are no differences in mortality rates between
the two drugs this analysis strategy provides unbiased estimates of
effectiveness of the NAM:s on morbidity. However, if there are differences
in mortality rates, these analyses are biased as it is more likely that we
observe a morbidity for the drug with lower mortality rates. If this is the
case, we will need to estimate bounds of potential effects as a sensitivity
analysis. We let all patients who die either have the morbidity or not (i.e.
PAIN =1 and SRE=1 and PAIN =0 and SRE=0). If the mortality with the NAM is
observed to be higher than without a NAM the first case provides the lower 
bound estimate of the effectiveness of the NAM while the second one provides
the upper bound on these two morbidity outcomes and vice versa if the
mortality rate is lower for the NAM:s.

\subsection{Sub-analyses}

Sub-analyses on these three outcomes will be conducted for the two groups
being prescribed NAM early and late after diagnosis. We estimate the the regression (%
\ref{OLS}) separately for those with less than medium DTP and longer or
equal medium DTP. 

\subsection{Sensitivity analysis}

While the design yields balanced observed covariates, this is an
observational study with the usual limitations in this context. 
In particular, one cannot discard the possibility that unobserved confounders
are not balanced. 

For this reason placebo regression will be estimated, that is we will estimate effect where we should not find effects given the design is valid. 

We will add data containing detailed information on patient's heath with regard to the prostate cancer from the national prostate cancer register (NPCR). The data is ordered and we will have access to it at the same time
as we have the mortality data. 

For the test we will use three covariates, that during discussions with specialists, are judged to be important confounders: PSA levels, Gleason score, and metastases at the time of prostate cancer diagnoses. These three pre-treatment covariates will be used as outcomes in the same regression as for the main analysis. If we find a statistical significant effect on these outcome this is signal of hidden bias in our analyses. 

With three pre-measured covariates we, as in the main analysis, adjust the significance level for the individual tests using Bonferroni correction based on a five percent overall level. This means that
the level for testing on each single outcome is 1.67 percent.

\subsection{Duration analysis}
We choose to model the prevalence of mortality as the outcomes as the analysis is straight forward using a WLS estimator. However, we will also exploratory analyze the effect on the duration using a discrete time Cox regression model. 

Let $Y_{im}=1$ if patient $i$ dies in month $m$ and let $p_{im}(\alpha_t,T_i)=\Pr (Y_{im}=1|\alpha_t,T_{i},Y_{im-1}=0)$ be the probability that this patient dies at month $m$ given survival up to this month. The discrete time Cox regression model is obtained by letting
\begin{equation*}
p_{im}(\mathbf{z}_{i},T_i)=1-\exp \left( -e^{\gamma_{m}+ \alpha_{t} + \tau T_i}\right),
\end{equation*}%
that is, the complementary log-log (CLL) specification. The parameter of interest is $\tau$ which is a constant shift in the baseline hazard determined by $\mathbf{\gamma}+\mathbf{\alpha}$, where $\mathbf{\gamma}=(\gamma _{1},...\gamma _{M-1})^{\prime }$ and $\mathbf{\alpha}= (\alpha_{1},..\alpha_{36})^{\prime}$, respectively. 

The parameters, $\mathbf{\theta =(\gamma }^{\prime },\mathbf{\alpha}^{\prime },\tau)^{\prime }$, are estimated using weights obtained from the entropy balancing scheme in a weighted maximum likelihood estimator (WMLE). The log likelihood to be maximized is:

\begin{equation}
\ell(\mathbf{\theta |}m,\mathbf{z})=\sum
\limits_{i=1}^{n}\sum_{m=1}^{m_{i}}Y_{im}\ln p_{im}(\alpha_t,T_i)+(1-Y_{im})\ln (1- p_{im}(\alpha_t,T_i)),
\end{equation}
where $m_i= M$ if the patient is alive at the censoring month $M\geq 24$.  

\section{Discussion}
\label{sec:Discussion}

The aim of this paper was to present a study protocol for an effectiveness
evaluation of antiandrogenic (NAM) medications (abiraterone acetate and
enzalutamide) in patients suffering from metastatic castrate resistant
prostate cancer (mCRPC) in contrast to standard of care. The study design takes stock in the Rubin Causal Model \citep{Rubin1974} and makes use of a weighting estimator defined using entropy balancing \citep{hainmueller2012entropy} in the design and estimation of the effects for the NAM treated on mortality, pain and skeleton related events 

The Dental and Pharmaceutical Benefit Agency (TLV) decided in June 2015 that 
NAM medication to be reimbursed when given to mCRPC patients. Before subsidizing
the drugs, NAM medication were not used in clinical practice. The evaluation
concerns their use in clinical practice from June 2015 to June 15 2018. We control for a large set of covariates before the diagnosis but also, importantly, a large set of covariates capturing the health progression in the period between the diagnosis and prescription (DTP). 

The fact that comparisons were not offered NAM:s mean explains why the patients were given two different treatments while at the same time having the same observed health. Thus, this fact is a valid argument for the common support assumption of the design.

The evaluation is restricted to NAM treated with a prostate cancer (PC) diagnosis 
in June 2012 to June 2015 with a DTP $\leq 36$. The comparison group is sampled among men that had a PC diagnosis between June 2008 and June 2010. The design takes into consideration that the comparisons cannot be
prescribed a NAM when measuring the outcomes, which was the reason to restrict the population of NAM treated to patients prescribed the drugs within 36 months. 

As the sampling of treated and comparisons is close in time the demography of the two groups are similar and they should have access to similar quality of health care. The sampling
scheme leaves us with 1,285 treated patients and a large group of 19,456
comparisons. As the comparison group is almost twenty times that of the
treatment group this enables us to find control individuals to create
balance between the groups in observed socioeconomic and health covariates
measure before diagnosis as well as among covariates measuring health
progression up to NAM treatment.

While the design yields balanced observed covariates, this is an
observational study with the usual limitations in this context. The
sensitivity to hidden bias will however be investigated estimating placebo regressions, that is, we will estimate effects were there should not be any effects given the design is valid.
These placebo regressions will be estimated using data from the national prostate cancer register (NPCR), that will be added, together with the morality data, after the publication of this pre-analysis plan. 

Generalizability of the results to other populations may not necessarily be granted, for instance, to patient with longer duration to be prescribed a NAM from diagnosis or to future populations who differ greatly in
characteristics than in this study.

\newpage
\section*{Appendix}

\begin{figure}[H]
\centerline{
\makebox[\textwidth][c]{\includegraphics[width=.8\textwidth]{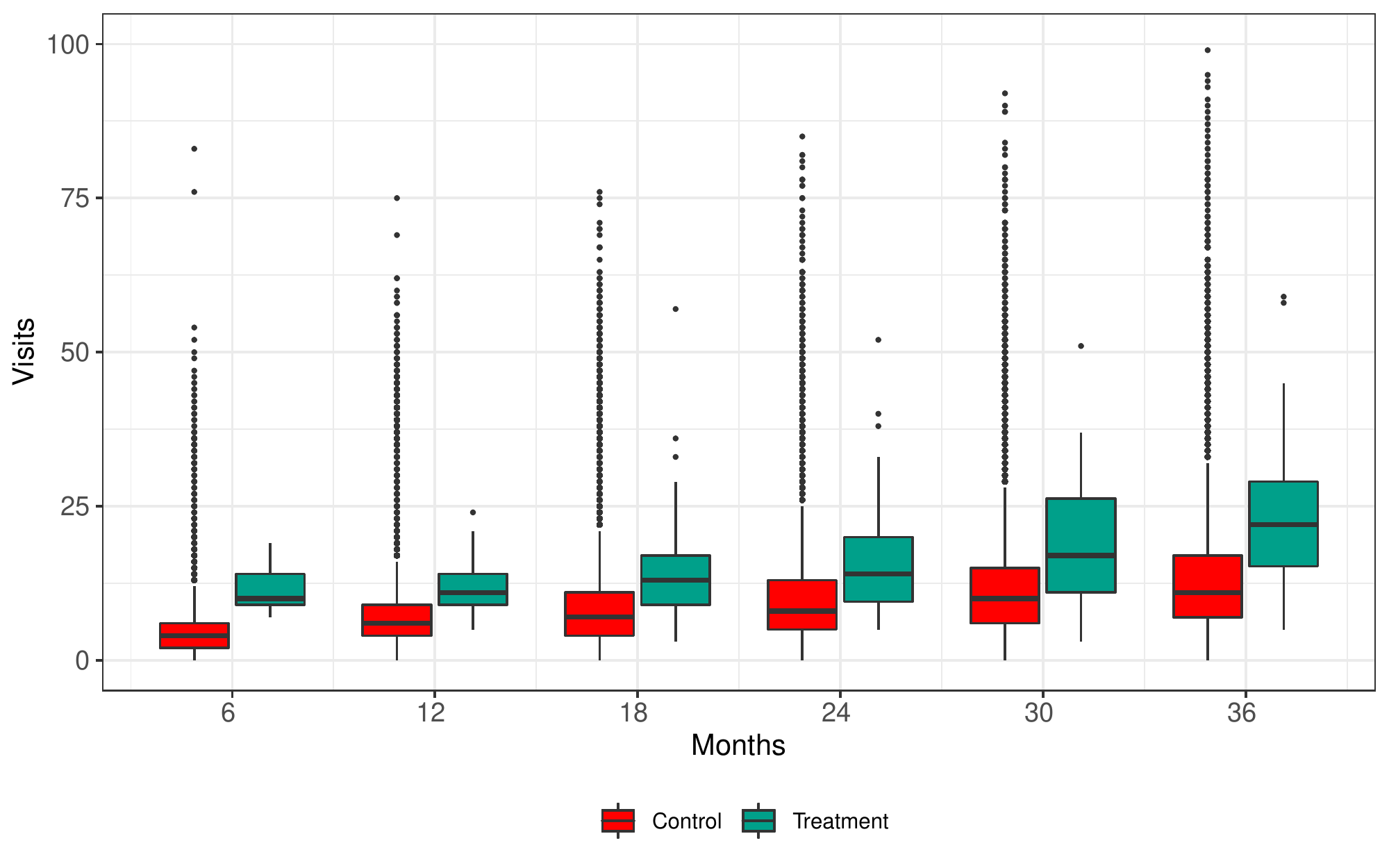}} 
}
\caption{Distribution of number of visits for those NAM treated 6, 12, 18, 24, 30 and 36 months after diagnosis and the corresponding comparison groups.}
\label{VISITS2}
\end{figure}

\begin{figure}[H]
\centerline{
\makebox[\textwidth][c]{\includegraphics[width=.8\textwidth]{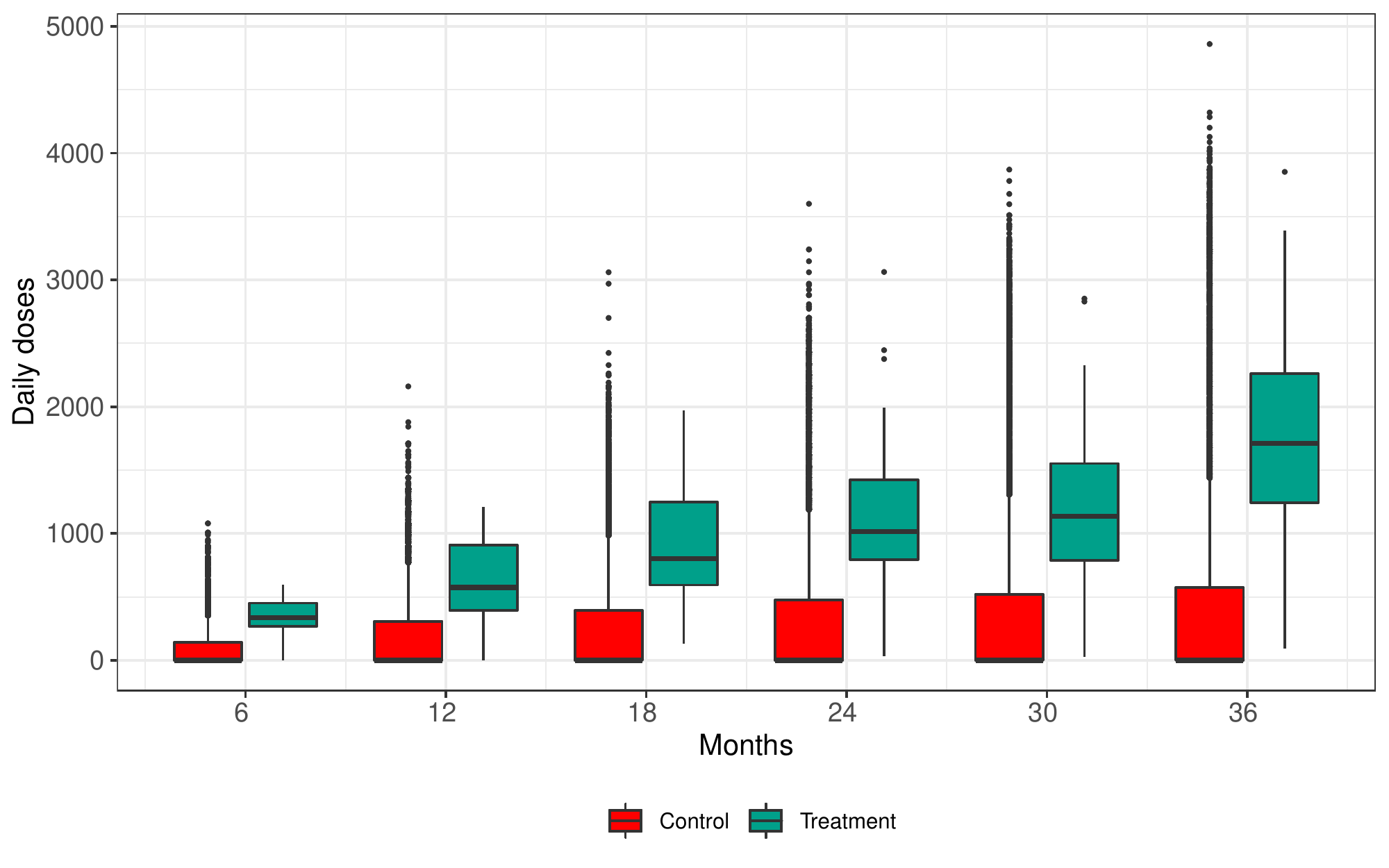}} 
}
\caption{Distribution of number of defined daily doses for those NAM treated 6, 12, 18, 24, 30 and 36 months after diagnosis and the corresponding comparison groups.}
\label{DAYS2}
\end{figure}

\newpage

\begin{table}[H]
\caption{Factor model with 5 factors, socioeconomic variables}{\footnotesize 
\label{Factormodel}
\centering
\begin{tabular}{lrrrrr}
\hline
& Factor1 & Factor2 & Factor3 & Factor4 & Factor5 \\ \hline
PrivPens\_1y & 0.96 &  &  &  &  \\ 
  AldPens\_1y & 0.96 &  &  &  &  \\ 
  AldPens & 0.95 &  &  &  &  \\ 
  PrivPens & 0.95 &  &  &  &  \\ 
  SumTjp & 0.95 &  &  &  &  \\ 
  AldPens\_2y & 0.92 &  &  &  &  \\ 
  SumTjp\_1y & 0.91 &  &  &  &  \\ 
  PrivPens\_2y & 0.82 &  &  &  &  \\ 
  SumTjp\_2y & 0.69 &  &  &  &  \\ 
  LoneInk\_2y & -0.26 &  &  & 0.22 & 0.63 \\ 
  LoneInk\_1y & -0.31 &  &  & 0.22 & 0.47 \\ 
  LoneInk & -0.39 &  &  & 0.21 & 0.21 \\ 
  SocBidrPers\_1y &  & 0.95 &  &  &  \\ 
  SocBidrFam\_1y &  & 0.95 &  &  &  \\ 
  SocBidrFam &  & 0.89 &  &  &  \\ 
  SocBidrPers &  & 0.89 &  &  &  \\ 
  SocBidrFam\_2y &  & 0.87 &  &  &  \\ 
  SocBidrPers\_2y &  & 0.86 &  &  &  \\ 
  ForTid\_1y &  &  & 0.95 &  &  \\ 
  ForTid &  &  & 0.90 &  &  \\ 
  SocInk\_1y &  &  & 0.89 &  &  \\ 
  ForTid\_2y &  &  & 0.87 &  &  \\ 
  SocInk\_2y &  &  & 0.84 &  &  \\ 
  SocInk &  &  & 0.78 &  &  \\ 
  SjukRe\_2y &  &  & 0.23 &  &  \\ 
  DispInk\_2y &  &  &  & 0.95 &  \\ 
  KapInk\_2y &  &  &  & 0.93 &  \\ 
  DispInkFam\_2y &  &  &  & 0.89 &  \\ 
  DispInk &  &  &  & 0.49 &  \\ 
  DispInkFam &  &  &  & 0.47 &  \\ 
  KapInk &  &  &  & 0.43 &  \\ 
  DispInkFam\_1y &  &  &  & 0.28 & 0.93 \\ 
  DispInk\_1y &  &  &  & 0.25 & 0.94 \\ 
  KapInk\_1y &  &  &  &  & 0.94 \\ 
  SjukRe &  &  &  &  &  \\ 
  SjukRe\_1y &  &  &  &  &  \\ 
  ArbLos &  &  &  &  &  \\ 
  ArbLos\_1y &  &  &  &  &  \\ 
  ArbLos\_2y &  &  &  &  &  \\ 
  InkFNetto &  &  &  &  &  \\ 
  InkFNetto\_1y &  &  &  &  &  \\ 
  InkFNetto\_2y &  &  &  &  &  \\ 
   \hline
\hline
SS loadings & 7.90 & 4.93 & 4.84 & 3.61 & 3.43 \\ 
  Proportion Var & 0.19 & 0.12 & 0.12 & 0.09 & 0.08 \\ 
  Cumulative Var & 0.19 & 0.31 & 0.42 & 0.51 & 0.59 \\ \hline
\end{tabular}
}
\end{table}

\begin{table}[ht]
\centering
\ContinuedFloat
\caption{All variables (* variable is included in factor analysis)}
\label{All_vars}
\begin{tabular}{llll}
\hline
& Variable name & Description & \\ \hline
 & LOPNR & ID number & \\ 
 & DDAT & Date of diagnosis & \\ 
& ALDER & Age at diagnosis & \\ 
& CIVIL & Marital status at diagnosis & \\
& Fodelseland\_EU28 & Born in the nordic countries or not & \\
  & UTBNFORGYMN & Educational level at  diagnosis: less than secondary school \\ 
 & UTBNGYMN & Educational level at diagnosis: secondary school \\ 
 & UTBNEFTERGYMN & Educational level at diagnosis: more than secondary school \\
 & MINDateF & Date of prescription ENZ/AA & \\ 
 & TREAT & Treatment $= 1$, Control $=0$ & \\ 
 & SCORE\_0 & Elixhouser score at diagnosis & \\ 
 & SCORE\_1 & Elixhouser score one month before $T=T^*=2$ & \\ 
 & $\vdots$ & $\vdots$ & \\ 
 & SCORE\_35 & Elixhouser score one month before $T=T^*=36$ & \\
 & VISITS\_0 & Number of visits at the month of diagnose & \\ 
 & VISITS\_1 & Number of visits 1 month after the diagnose & \\ 
 & $\vdots$ & $\vdots$ & \\ 
 & VISITS\_35 & Number of visits 35 months after diagnose \\ 
 & METAS\_0 & Presence of metastases at the month of diagnose & \\ 
 & METAS\_1 & Presence of metastases up until 1 month after the diagnose & 
\\ 
 & $\vdots$ & $\vdots$ & \\ 
  & METAS\_35 & Presence of metastases up until 35 months after the diagnose & \\ 
 & MEDS\_DAYS\_1 & Days with prescribed ADT between diagnosis and $T=
T^*=2$ & \\ 
& $\vdots$ & $\vdots$ & \\ 
 & MEDS\_DAYS\_35 & Days with prescribed ADT between diagnosis and $T=
T^*=36$  & \\ 
 & VISITS\_BFD\_1 & Number of visits 1 month before diagnosis & \\ 
 & VISITS\_BFD\_6 & Number of visits 6 months before diagnosis & \\ 
 & VISITS\_BFD\_12 & Number of visits 612 months before diagnosis & \\ 
 & VISITS\_BFD\_60 & Number of visits 60 months before diagnosis & \\ 
 & SUM\_DAYS & Sum of inpatient care days after diagnosis up until $T=
T^*$ & 
\\ 
 & BIKA\_OUT\_1 & The patient is prescribed (1) only bikalutamide, & \\ 
& & (2) Bikalutamide+GnRH or (3) nothing up until  $T=T^*=2$ & \\ 
 & $\vdots$ & $\vdots$ & \\ 
 & BIKA\_OUT\_35 & The patient is prescribed (1) only bikalutamide, & \\ 
& & (2) Bikalutamide+GnRH or (3) nothing up until  $T=T^*=35$ & \\  
 & SUM\_VIS\_METAS & Months with visceral metastases before $T$ or $T^*$ & 
\\ 
 & SUM\_SK\_METAS & Months with skeleton metastases (C975) before $T$ or $%
T^*$ & \\ \hline
\end{tabular}%
\end{table}

\begin{table}[ht]
\centering
\ContinuedFloat
\caption{All variables (* variable is included in factor analysis) cont.}
\label{All_vars}
\begin{tabular}{llll}
\hline
& Variable name & Description & \\ \hline
* & LoneInk & Wage income at diagnosis \\ 
* & InkFNetto & Income from business at diagnosis \\ 
* & KapInk & Capital income at diagnosis \\ 	
* & DispInk & Disposable income at diagnosis \\ 
* & DispInkFam & Family disposable income at diagnosis \\ 
* & LoneInk\_1y & Wage income one year before diagnosis \\ 
* & InkFNetto\_1y & Income from business one year before diagnosis \\ 
* & KapInk\_1y & Capital income one year before diagnosis\\ 
*  & DispInk\_1y & Disposable income one year before diagnosis\\ 
* & DispInkFam\_1y & Family disposable income one year before diagnosis\\
* & LoneInk\_2y & Wage income two years before diagnosis \\ 
* & InkFNetto\_2y & Income from business two years before diagnosis \\ 
* & KapInk\_2y & Capital income two years before diagnosis\\ 
* & DispInk\_2y & Disposable two years before diagnosis\\ 
* & DispInkFam\_2y & Family disposable income two years before diagnosis\\
* & SjukRe & Sickness compensation at diagnosis \\ 
* & ArbLos & Unemployment benefits at diagnosis \\ 
* & ForTid & Early retirement benefit at diagnosis \\ 
* & SocInk & Social security benefits at diagnosis \\ 
* & SocBidrPersF & Social security benefits at diagnosis \\ 
* & SocBidrFam & Social security benefits of the family at diagnosis \\ 
* & SjukRe\_1y & Sickness compensation one year before diagnosis \\ 
* & ArbLos\_1y & Unemployment benefits one year before diagnosis \\ 
* & ForTid\_1y & Early retirement benefit one year before diagnosis \\ 
* & SocInk\_1y & Social security benefits one year before diagnosis \\
* & SocBidrPersF\_1y & Social security benefits one year before diagnosis \\ 
* & SocBidrFam\_1y & Social security benefits of the family one year before diagnosis \\ 
* & SjukRe\_2y & Sickness compensation two years before diagnosis \\ 
* & ArbLos\_2y & Unemployment benefits two years before diagnosis \\ 
* & ForTid\_2y & Early retirement benefit two years before diagnosis \\ 
* & SocInk\_2y & Social security benefits two years before diagnosis \\ 
* & SocBidrPersF\_2y & Social security benefits two years before diagnosis \\ 
* & SocBidrFam\_2y & Social security benefits of the family two years before diagnosis \\ 
* & AldPens & Old-age pensions at diagnosis \\ 
* & SumTjP & Occupational pensions one year before diagnosis \\ 
* & PrivPens & Private pensions two years before diagnosis \\ 
* & AldPens\_1y & Old-age pensions at diagnosis \\ 
* & SumTjP\_1y & Occupational pensions one year before diagnosis\\ 
* & PrivPens\_1y & Private pensions two years before diagnosis \\ 
* & AldPens\_2y & Old-age pensions at diagnosis \\ 
* & SumTjP\_2y & Occupational pensions one year before diagnosis\\ 
* & PrivPens\_2y & Private pensions two years before diagnosis \\ \hline
\end{tabular}%
\end{table}

\bibliographystyle{agsm}
\bibliography{Refs2}

@article{Imbens_Wooldridge_2009,
 Author = {Imbens, Guido W. and Wooldridge, Jeffrey M.},
 Title = {Recent Developments in the Econometrics of Program Evaluation},
 Journal = {Journal of Economic Literature},
 Volume = {47},
 Number = {1},
 Year = {2009},
 Month = {March},
 Pages = {5-86}
}

@inproceedings{Eicker_1967,
  title={{Limit theorems for regressions with unequal and dependent errors}},
  author={Eicker, F.},
  booktitle={Proceedings of the Fifth Berkeley Symposium on Mathematical Statistics and Probability},
  volume={I},
  pages={59-82},
  publisher={University California Press, Berkeley, CA},
  year={1967}
}

@inproceedings{Huber_1967,
  title={{The behavior of maximum likelihood estimates under nonstandard conditions}},
  author={Huber, P. J.},
  booktitle={Proceedings of the Fifth Berkeley Symposium on Mathematical Statistics and Probability},
  volume={I},
  pages={221-233},
  publisher={University California Press, Berkeley, CA},
  year={1967}
}

@article{White_1980,
  title={{Using least squares to approximate unknown regression functions}},
  author={White, H.},
  journal={International Economic Revi},
  volume={21},
  number=1,
  pages={149-170},
  year={1980}
}

@article{Schuklenk_2003,
author = { Udo   Schuklenk },
title = {AIDS: Bioethics and public policy},
journal = {New Review of Bioethics},
volume = {1},
number = {1},
pages = {127-144},
year  = {2003},
publisher = {Routledge},
doi = {10.1080/1740028032000131477},
    note ={PMID: 15706680},

URL = { 
        https://doi.org/10.1080/1740028032000131477
    
},
eprint = { 
        https://doi.org/10.1080/1740028032000131477
    
}

}

@article{Teira_2013,
author = {David Teira},
title ={Blinding and the Non-interference Assumption in Medical and Social Trials},
journal = {Philosophy of the Social Sciences},
volume = {43},
number = {3},
pages = {358-372},
year = {2013},
doi = {10.1177/0048393113488871},

URL = { 
        https://doi.org/10.1177/0048393113488871
    
},
eprint = { 
        https://doi.org/10.1177/0048393113488871
    
}
}

@misc{Johansson_et_al_2021a,
author={Johansson, Per and Joneus, Paulina and Langenskiold, Sophie},
year = {2021},
title ={Study protocol for a comparative effectiveness evaluation of abiraterone against enzalutamide},
note= {Mimeo Uppsla university January, 2021}
}

@article{Neyman1923,
  title={edited and translated by {D}orota {M}. {D}abrowska and {T}errence {P}. {S}peed (1990). {O}n the Application of Probability Theory to Agricultural Experiments. {E}ssay on Principles. {S}ection 9},
  author={Neyman, Jerzy},
  journal={Statistical Science},
  volume={5},
  number={4},
  pages={465--472},
  year={1923}
}

@article{Rubin1974,
  title={Estimating causal effects of treatments in randomized and nonrandomized studies.},
  author={Rubin, Donald B},
  journal={Journal of educational Psychology},
  volume={66},
  number={5},
  pages={688},
  year={1974},
  publisher={American Psychological Association}
}

@article{holland1986,
  title={Statistics and causal inference},
  author={Holland, Paul W},
  journal={Journal of the American statistical Association},
  volume={81},
  number={396},
  pages={945--960},
  year={1986},
  publisher={Taylor \& Francis}
}

@article{rawla19,
  title={Epidemiology of prostate cancer},
  author={Rawla, Prashanth},
  journal={World journal of oncology},
  volume={10},
  number={2},
  pages={63},
  year={2019},
  publisher={Elmer Press}
}

@article{hainmueller2012entropy,
  title={Entropy balancing for causal effects: A multivariate reweighting method to produce balanced samples in observational studies},
  author={Hainmueller, Jens},
  journal={Political analysis},
  pages={25--46},
  year={2012},
  publisher={JSTOR}
}

\end{document}